\documentclass[longauth]{aa}  

\usepackage[varg]{txfonts}
\usepackage{graphicx}
\usepackage{amsmath}
\usepackage{amssymb}
\usepackage{color}
\usepackage{natbib}
\bibpunct{(}{)}{;}{a}{}{,} 
\usepackage{url}
\usepackage{todonotes}
\usepackage{multirow}
\usepackage{threeparttable}      
\usepackage{booktabs} 
\usepackage{soul}
\usepackage{float}
\usepackage[colorlinks=true]{hyperref}
\hypersetup{colorlinks=true,citecolor=blue}

\begin{document}

\title{Quantifying thermal water dissociation  in the dayside photosphere of WASP-121\,b using NIRPS}

\author{
Luc Bazinet\inst{1,*},
Romain Allart\inst{1},
Bj\"orn Benneke\inst{2,1},
Stefan Pelletier\inst{3,1},
Joost P. Wardenier\inst{1},
Neil J. Cook\inst{1},
Thierry Forveille\inst{4},
Louise D. Nielsen\inst{3,5,6},
Khaled Al Moulla\inst{7,3},
\'Etienne Artigau\inst{1,8}
% ...
}

\institute{
\inst{1}Institut Trottier de recherche sur les exoplan\`etes, D\'epartement de Physique, Universit\'e de Montr\'eal, Montr\'eal, Qu\'ebec, Canada\\
\inst{2}Department of Earth, Planetary, and Space Sciences, University of California, Los Angeles, CA 90095, USA\\
\inst{3}Observatoire de Gen\`eve, D\'epartement d’Astronomie, Universit\'e de Gen\`eve, Chemin Pegasi 51, 1290 Versoix, Switzerland\\
\inst{4}Univ. Grenoble Alpes, CNRS, IPAG, F-38000 Grenoble, France\\
\inst{5}European Southern Observatory (ESO), Karl-Schwarzschild-Str. 2, 85748 Garching bei M\"unchen, Germany\\
\inst{6}University Observatory, Faculty of Physics, Ludwig-Maximilians-Universit\"at M\"unchen, Scheinerstr. 1, 81679 Munich, Germany\\
\inst{7}Instituto de Astrof\'isica e Ci\^encias do Espa\c{c}o, Universidade do Porto, CAUP, Rua das Estrelas, 4150-762 Porto, Portugal\\
\inst{8}Observatoire du Mont-M\'egantic, Qu\'ebec, Canada\\
\inst{9}Departamento de F\'isica e Astronomia, Faculdade de Ci\^encias, Universidade do Porto, Rua do Campo Alegre, 4169-007 Porto, Portugal\\
\inst{10}Department of Physics, University of Toronto, Toronto, ON M5S 3H4, Canada\\
\inst{11}Departamento de F\'isica Te\'orica e Experimental, Universidade Federal do Rio Grande do Norte, Campus Universit\'ario, Natal, RN, 59072-970, Brazil\\
\inst{12}Department of Physics \& Astronomy, McMaster University, 1280 Main St W, Hamilton, ON, L8S 4L8, Canada\\
\inst{13}Department of Physics, McGill University, 3600 rue University, Montr\'eal, QC, H3A 2T8, Canada\\
\inst{14}Department of Earth \& Planetary Sciences, McGill University, 3450 rue University, Montr\'eal, QC, H3A 0E8, Canada\\
\inst{15}Departamento de F\'isica, Universidade Federal do Cear\'a, Caixa Postal 6030, Campus do Pici, Fortaleza, Brazil\\
\inst{16}Centre Vie dans l’Univers, Facult\'e des sciences de l’Universit\'e de Gen\`eve, Quai Ernest-Ansermet 30, 1205 Geneva, Switzerland\\
\inst{17}Instituto de Astrof\'isica de Canarias (IAC), Calle V\'ia L\'actea s/n, 38205 La Laguna, Tenerife, Spain\\
\inst{18}Departamento de Astrof\'isica, Universidad de La Laguna (ULL), 38206 La Laguna, Tenerife, Spain\\
\inst{19}Space Research and Planetary Sciences, Physics Institute, University of Bern, Gesellschaftsstrasse 6, 3012 Bern, Switzerland\\
\inst{20}Consejo Superior de Investigaciones Cient\'ificas (CSIC), E-28006 Madrid, Spain\\
\inst{21}Bishop's Univeristy, Dept of Physics and Astronomy, Johnson-104E, 2600 College Street, Sherbrooke, QC, Canada, J1M 1Z7\\
\inst{22}Department of Physics, Engineering Physics, and Astronomy, Queen’s University, 99 University Avenue, Kingston, ON K7L 3N6, Canada\\
\inst{23}Department of Physics and Space Science, Royal Military College of Canada, 13 General Crerar Cres., Kingston, ON K7P 2M3\\
\inst{24}Instituto de Astrof\'isica e Ci\^encias do Espa\c{c}o, Faculdade de Ci\^encias da Universidade de Lisboa, Campo Grande, 1749-016 Lisboa, Portugal\\
\inst{25}Departamento de F\'isica da Faculdade de Ci\^encias da Universidade de Lisboa, Edif\'icio C8, 1749-016 Lisboa, Portugal\\
\inst{26}Centre of Optics, Photonics and Lasers, Universit\'e Laval, Qu\'ebec, Canada\\
\inst{27}Herzberg Astronomy and Astrophysics Research Centre, National Research Council of Canada\\
\inst{28}Center for Space and Habitability, University of Bern, Gesellschaftsstrasse 6, 3012 Bern, Switzerland\\
\inst{29}Aix Marseille Univ, CNRS, CNES, LAM, Marseille, France\\
\inst{30}European Southern Observatory (ESO), Av. Alonso de Cordova 3107,  Casilla 19001, Santiago de Chile, Chile\\
\inst{31}Plan\'etarium de Montr\'eal, Espace pour la Vie, 4801 av. Pierre-de Coubertin, Montr\'eal, Qu\'ebec, Canada\\
\inst{32}Lund Observatory, Division of Astrophysics, Department of Physics, Lund University, Box 118, 221 00 Lund, Sweden\\
\inst{33}York University, 4700 Keele St, North York, ON M3J 1P3\\
\inst{34}University of British Columbia, 2329 West Mall, Vancouver, BC, Canada, V6T 1Z4\\
\inst{35}Western University, Department of Physics \& Astronomy and Institute for Earth and Space Exploration, 1151 Richmond Street, London, ON N6A 3K7, Canada\\
\inst{36}Light Bridges S.L., Observatorio del Teide, Carretera del Observatorio, s/n Guimar, 38500, Tenerife, Canarias, Spain\\
\inst{37}Department of Astronomy \& Astrophysics, University of Chicago, 5640 South Ellis Avenue, Chicago, IL 60637, USA\\
\inst{38}Laboratoire Lagrange, Observatoire de la C\^ote d’Azur, CNRS, Universit\'e C\^ote d’Azur, Nice, France\\
\inst{*}\email{luc.bazinet@umontreal.ca}
}

\date{Received 10 January 2025; accepted 4 August 2025}

\abstract{
The intense stellar irradiation of ultra-hot Jupiters results in some of the most extreme atmospheric environments in the planetary regime. On their daysides, temperatures can be sufficiently high for key atmospheric constituents to thermally dissociate into simpler molecular species and atoms. This dissociation drastically changes the atmospheric opacities and, in turn, critically alters the temperature structure, atmospheric dynamics, and day-night heat transport. To this date, however, simultaneous detections of the dissociating species and their thermally dissociation products in exoplanet atmospheres have remained rare. Here we present the simultaneous detections of H$_2$O and its thermally dissociation product OH on the dayside of the ultra-hot Jupiter WASP-121\,b based on high-resolution emission spectroscopy with the recently commissioned \textit{Near InfraRed Planet Searcher} (NIRPS). We retrieve a photospheric abundance ratio of log$_{10}$(OH/H$_2$O) $= -0.15\pm{0.20}$ indicating that there is about as much OH as H$_2$O at photospheric pressures, which confirms predictions from chemical equilibrium models. We compare the dissociation on WASP-121\,b with other ultra-hot Jupiters and show that a trend in agreement with equilibrium models arises. We also discuss an apparent velocity shift of $4.79^{+0.93}_{-0.97}\,$km$\,$s$^{-1}$ in the H$_2$O signal, which is not reproduced by current global circulation models. Finally, in addition to H$_2$O and OH, the NIRPS data reveal evidence of Fe and Mg, from which we infer a Fe/Mg ratio consistent with the solar and host star ratios.
Our results demonstrate that NIRPS can be an excellent instrument to obtain simultaneous measurements of refractory and volatile molecular species, paving the way for many future studies on the atmospheric composition, chemistry, and the formation history of close-in exoplanets.}

\titlerunning{Thermal water dissociation on WASP-121\,b}
\authorrunning{L. Bazinet et al.}
\maketitle
   
\section{Introduction} \label{sec:intro}

Ultra-hot Jupiters (UHJs) are extreme worlds. Their elevated equilibrium temperature (T$_{\mathrm{eq}}$ $>$ 2200\,K) introduces atmospheric dynamics unseen on other planets. Among hot Jupiters, they are uniquely characterised by their dayside temperature profile: they have a pressure layer where the temperature increases with increasing altitude (and decreasing pressure), a feature called a temperature inversion \citep{fortney_unified_2008, madhusudhan_inference_2010, gandhi_new_2019}. Temperature inversions on UHJs are not only predicted by theory but have also been observed \citep[e.g.][]{evans_ultrahot_2017, mikal-evans_emission_2019, coulombe_broadband_2023, mikal-evans_diurnal_2022, evans-soma_sio_2025}.
An atmospheric inversion creates emission lines in the dayside spectra, instead of absorption features seen on planets with a non-inverted profile.
Furthermore, some molecular species that are present in great quantity in the colder hot Jupiters are expected to be thermally dissociated in UHJs \citep{parmentier_thermal_2018}.
Notably, H$_2$O is prone to thermal dissociation, giving rise to significant amounts of hydroxyl (OH) and atomic hydrogen (H) \citep{parmentier_thermal_2018}.
On cooler planets, the temperature is not high enough for water to dissociate. Therefore, the relative amounts of OH and H$_2$O present in the atmosphere can be a good tracer of the temperature\citep{kitzmann_peculiar_2018}.

OH was detected for the first time in the emission spectrum of WASP-33\,b (T$_{\mathrm{eq}} \sim 2800$\,K) \citep{nugroho_first_2021, collier_cameron_line-profile_2010}.
Following this, \citet{finnerty_keck_2023} were able to reproduce their detection and performed retrievals to constrain the atmospheric abundance of OH, H$_2$O and other prominent species. They concluded that the OH abundance was two orders of magnitude higher than that of H$_2$O.
\citet{brogi_roasting_2023} were able to retrieve OH on the colder UHJ WASP-18\,b (T$_{\mathrm{eq}} \sim 2500$\,K) \citep{hellier_orbital_2009}, finding an abundance marginally higher than that of H$_2$O.
OH was also detected on WASP-76\,b (T$_{\mathrm{eq}} \sim 2200$\,K) \citep{west_three_2016} using transit data from CARMENES\footnote{Calar Alto high-Resolution search for M dwarf with Exoearths with Near-infrared and optical \'{E}chelle Spectrograph} \citep{landman_detection_2021, gandhi_revealing_2024}.
\citet{mansfield_metallicity_2024} used IGRINS\footnote{Immersion GRating INfrared Spectrograph} transit data of WASP-76\,b to detect OH and recover its abundance to be comparable but slightly lower than that of H$_2$O.
More recently, OH was found on the dayside of WASP-121\,b \citep{delrez_wasp-121_2016} using IGRINS \citep{smith_roasting_2024}.

With its ability to resolve individual spectral lines, high-resolution spectroscopy is an excellent technique to uncover the dynamics in atmospheres of planets \citep[e.g.][]{snellen_orbital_2010, louden_spatially_2015, brogi_rotation_2016, ehrenreich_nightside_2020, prinoth_titanium_2022, nortmann_crires_2024}.
In particular, Doppler shifts of molecular lines in transmission data can be indicative of day-night circulation \citep[e.g.][]{ehrenreich_nightside_2020, seidel_into_2021, wardenier_decomposing_2021, bello-arufe_mining_2022, seidel_detection_2023}.
Meanwhile, different velocity shifts between detected species in dayside emission data can be caused by differences in probed pressure levels, or an inhomogeneous spatial distribution of the chemical composition across the photosphere \citep[e.g.][]{cont_detection_2021, brogi_roasting_2023}.
Since hot Jupiters are expected to be in synchronous rotation, they have a permanent dayside where the irradiation of the star is always present, and a night side which never sees the light of the host star.
This day-to-night contrast in irradiation can lead to temperature differences between the day and night sides of sometimes more than 1500\,K \citep{dang_comprehensive_2024}, which can cause large longitudinal variations in atmospheric chemistry.
This important difference in energy between the two hemispheres and the rotation of the planet are responsible for dynamical effects such as the formation of a superrotating equatorial jet \citep[e.g.][]{showman_atmospheric_2009}.

Global circulation models (GCMs) were developed to better understand three-dimensional dynamic effects present on hot Jupiters \citep[e.g.][]{showman_atmospheric_2009, rauscher_general_2012}. 
GCMs simulate the 3D climate of the planet using the primitive equations of meteorology and a radiative-transfer prescription. External energy sources, such as the heating from the host star, are considered in the models. 
This will naturally give rise to atmospheric effects such as winds and longitude-dependent temperature profiles \citep[e.g.][]{harada_signatures_2021, malsky_modeling_2021}.
The effects of H$_2$ dissociation/recombination \citep[e.g.][]{bell_increased_2018, tan_atmospheric_2019, roth_pseudo-2d_2021} and magnetic drag \citep[e.g.][]{rauscher_three-dimensional_2013, beltz_exploring_2021, beltz_magnetic_2022} on the heat transport can also be included to obtain a more realistic view of a planet.
GCMs were successfully used to explain dynamical features that cannot be explained using 1D models \citep[e.g.][]{wardenier_modelling_2023, wardenier_phase-resolving_2024, kesseli_up_2024}.
With the advent of high-resolution spectroscopy at high signal-to-noise ratio, the dynamics of the atmosphere are observed in greater detail than ever before and the use of GCMs is sometimes necessary to properly interpret the results.

WASP-121\,b is a well-studied UHJ (T$_\mathrm{eq}$ $\sim$ 2350\,K) on a 30-hour orbit \citep{delrez_wasp-121_2016} of a relatively bright ($J$=9.6) F6V star. With its large radius \citep[R$_p$ = $1.753 \pm{0.036}$ R$_{\mathrm{Jup}}$,][]{bourrier_hot_2020}, low gravity, and its extreme irradiation, WASP-121\,b is an excellent target for both transmission and emission spectroscopy. 
In transmission, several chemical species were discovered using high resolution spectroscopy, including Fe \citep{gibson_detection_2020}, Cr, V and Fe$^{+}$ \citep{ben-yami_neutral_2020}, Mg, Ca, Ni \citep{hoeijmakers_hot_2020}, Na, K, Li, Ca$^{+}$ \citep{borsa_atmospheric_2021}, Sc$^{+}$ \citep{merritt_inventory_2021}, Co, Sr$^{+}$, Ba$^{+}$, and Mn \citep{azevedo_silva_detection_2022}.
Furthermore, H$_2$O and CO have been detected at infrared wavelengths using IGRINS \citep{wardenier_phase-resolving_2024}.

Compared to transmission, high-resolution studies of WASP-121\,b in emission are more sparse.
\citet{hoeijmakers_mantis_2024} used the ESPRESSO\footnote{Echelle SPectrograph for Rocky Exoplanets and Stable Spectroscopic Observations} spectrograph to uncover the presence of Ca, V, Cr, Mn, Fe, Co and Ni.
\citet{pelletier_crires_2024} combined the ESPRESSO dataset with infrared observations taken with CRIRES+\footnote{The CRyogenic InfraRed Echelle Spectrograph upgrade project} $K$-band data to further obtain detections of H$_2$O and CO on the dayside of WASP-121\,b.
Finally, \citet{smith_roasting_2024} found H$_2$O, CO and OH using IGRINS.

\begin{table*}[h]
\centering
\begin{threeparttable}
    \caption{Properties of the planet WASP-121\,b and its stellar host.}
    \begin{tabular}{lccc}
        \hline \hline
        Parameter & Symbol [unit] & Value & Reference \\
        \hline
        \noalign{\smallskip}
        \textbf{Star} &&& \\
        \noalign{\smallskip}
        Spectral type           &                                   & F6V                           & \citet{delrez_wasp-121_2016} \\
        Effective temperature   & T$_\mathrm{eff}$ [K]              & 6459\,$\pm$\,140              & \citet{delrez_wasp-121_2016} \\
        Radius                  & R$_*$ [R$_\odot$]                 & 1.46\,$\pm$\,0.03             & \citet{delrez_wasp-121_2016} \\
        Mass                    & M$_*$ [M$_\odot$]                 & 1.36\,$^{+0.07}_{-0.08}$      & \citet{delrez_wasp-121_2016} \\
        RV semi-amplitude       & K$_*$ [m\,s$^{-1}$]                       & 177.0\,$^{+8.5}_{-8.1}$       & \citet{bourrier_hot_2020} \\
        Systemic velocity       & V$_\mathrm{sys}$ [km\,s$^{-1}$]           & 38.12                         & Line-by-line \citep{artigau_line-by-line_2022}  \\
        \hline
        \noalign{\smallskip}
        \textbf{Planet} &&& \\
        \noalign{\smallskip}
        Radius                  & R$_\mathrm{p}$ [R$_\mathrm{Jup}$] & 1.753\,$\pm$\,0.036           & \citet{bourrier_hot_2020} \\ 
        Mass                    & M$_\mathrm{p}$ [M$_\mathrm{Jup}$] & 1.157\,$\pm$\,0.070           & \citet{bourrier_hot_2020} \\
        Planet surface gravity  & g$_\mathrm{p}$ [m\,s$^{-2}$]          & 9.76                          & \citet{bourrier_hot_2020}\tnote{1} \\
        Inclination             & i [deg]                           & 88.49\,$\pm$\,0.16            & \citet{bourrier_hot_2020} \\
        Semi-major axis         & a [AU]                            & 0.02596\,$^{+0.00043}_{-0.00063}$  & \citet{bourrier_hot_2020} \\
        Orbital period          & P [days]                          & 1.27492504\,$^{+1.5 \times 10^{-7}}_{-1.4 \times 10^{-7}}$ & \citet{bourrier_hot_2020} \\
        RV semi-amplitude       & K$_\mathrm{p}$ [km\,s$^{-1}$]             & 216.8\,$\pm\,4.5$          & \citet{bourrier_hot_2020}\tnote{1} \\
        Linear velocity at equator & V$_{\mathrm{rot}}$ [km\,s$^{-1}$] & 6.99\,$\pm$\,0.14 & \citet{bourrier_hot_2020}\tnote{1} \\
        Equilibrium temperature & T$_\mathrm{eq}$ [K]               & 2358\,$\pm$\,52             & \citet{delrez_wasp-121_2016} \\
         \hline
    \end{tabular}
\begin{tablenotes}
 \item[1] Derived parameter
\end{tablenotes}
\label{tab:properties}
\end{threeparttable}
\end{table*}

In this work, we analyse observations of the dayside emission of WASP-121\,b taken with the Near InfraRed Planet Searcher (NIRPS) and find evidence of thermal dissociation and dynamical effects in its atmosphere.
In Sect. \ref{sec:obs}, we present the observations used in this analysis and the NIRPS spectrograph. In Sect. \ref{sec:data}, an overview of the data processing is provided. The modelling and cross-correlation results are presented in Sect. \ref{sec:model}. The retrieval setup and associated results are shown in Sect. \ref{sec:retrieval}. The inferred atmospheric composition and dynamics are discussed in Sect. \ref{sec:discussion}. We conclude this manuscript in Sect. \ref{sec:conclusion}.

\section{Observations} \label{sec:obs}

\begin{figure*}
    \centering
    \includegraphics[width=\linewidth]{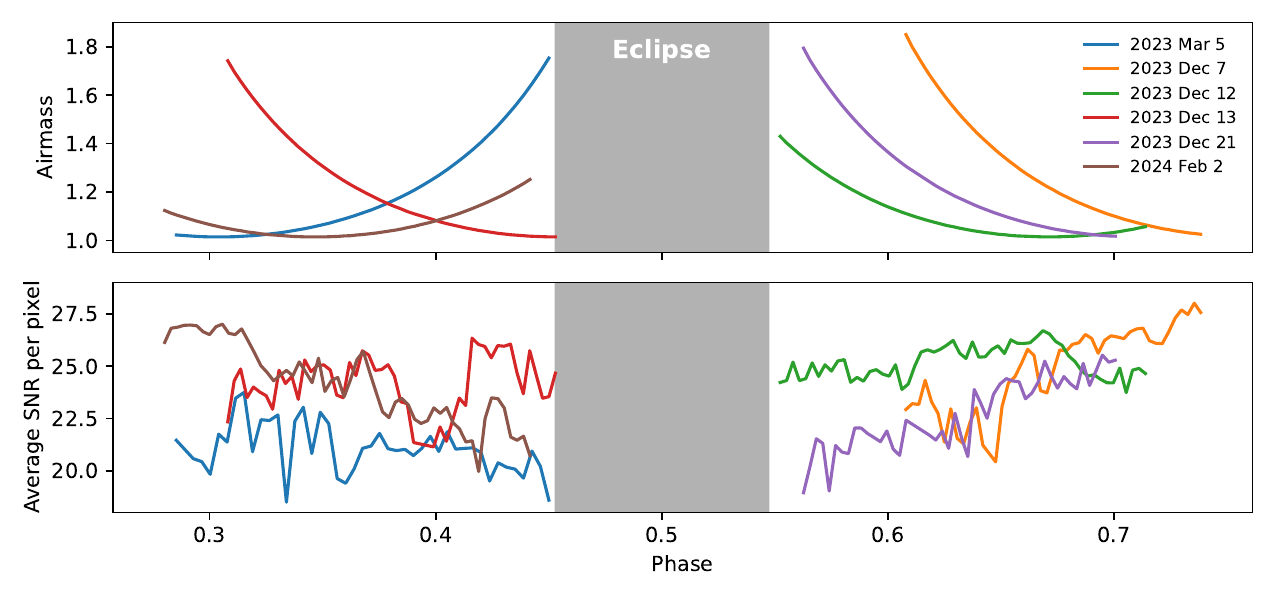}
    \caption{Summary of the observations of WASP-121\,b. The top panel is the airmass as a function of phase for all observations. The bottom panel shows the S/N per pixel throughout the observations. The grey shaded area centred around phase 0.5 represents the phases where WASP-121\,b is eclipsed by its host star.
    Of the six observations, three of them were obtained during the pre-eclipse phases (phases $<$ 0.5), while the other three were of the post-eclipse phases (phases $>$ 0.5).
    }
    \label{fig:obs}
\end{figure*}

We observed the dayside of WASP-121\,b using the Near InfraRed Planet Searcher  \citep[NIRPS;][Bouchy et al. submitted]{bouchy_near-infrared_2017, wildi_first_2022, artigau_nirps_2024} spectrograph.
NIRPS is the new infrared spectrograph on the ESO 3.6-m telescope at La Silla Observatory in Chile. 
With its wavelength coverage between 0.98 and 1.8$\,\mu$m, NIRPS covers the \textit{Y}, \textit{J} and \textit{H} bands simultaneously.
It is a fibre-fed, adaptive-optics-assisted, and ultra-stable spectrograph, making it an excellent tool for high-precision radial velocity measurements and atmospheric characterisation of exoplanets.
Two fibres are available for observations: the High-Accuracy (HA) and High-Efficiency (HE) modes. The HA mode uses a 0.4$^{\prime\prime}$ fibre with a median spectral resolution of 88,000. The HE mode uses a bigger 0.9$^{\prime\prime}$ fibre, at the cost of a lower median spectral resolution of 75,200 (Bouchy et al. submitted).
All WASP-121\,b observations were taken with the HE mode, with the exception that the 2023 March 5 observations which were taken with the HA fibre.

From radial velocity measurements, NIRPS was evaluated to be extremely stable to below 1.5\,m\,s$^{-1}$ (Bouchy et al. accepted). This is significantly lower than the required precision for atmospheric characterisation of close-in giant planets, where the planet's radial velocity generally vary in the order of tens to hundreds of km\,s$^{-1}$ within a given time series of a few to several hours.

NIRPS complements the optical light instrument High Accuracy Radial velocity Planet Searcher (HARPS), installed on the same telescope \citep{pepe_harps_2002, mayor_setting_2003}.
While both instruments can be used simultaneously during an observation, with HARPS adding its 0.38 to 0.69\,$\mu$m wavelength range to the NIRPS infrared coverage, we focus the analysis presented in this manuscript on the NIRPS data.

Along with precise radial velocity measurements (Bouchy et al. submitted, Suárez Mascareño et al. in prep), NIRPS has so far been used to characterise the atmospheric physics and chemistry of several hot gas giants.
Helium was found on WASP-69\,b (Allart et al. submitted) and H$_2$O was detected in transmission on WASP-127\,b (Bouchy et al. submitted).
Furthermore, two transits of WASP-189\,b were observed with NIRPS and HARPS simultaneously. 
Although Fe was detected in the HARPS dataset, it was not found in the NIRPS dataset. This non-detection is attributed to the strong opacity of H$^-$ in the infrared (Vaulato et al. submitted).

We observed WASP-121\,b for a total of 30 hours spread across six nights using NIRPS as part of the guaranteed time observations (Program IDs 60.A-9109 and 112.25P3.001, PI: Bouchy).
The dataset is composed of three pre-eclipse observations taken on the nights of 2023 March 5, 2023 December 13 and 2024 February 2. The other three observations covered the post-eclipse phases, taken on the nights of 2023 December 7, 12 and 21 (Fig. \ref{fig:obs}). The average S/N per pixel across the full wavelength domain was between 18 and 28 for all observations.
The exposure times were 300\,s with the exception of 2023 March 5 where the exposure times were set to 400\,s due to worse seeing conditions. The observation sequences are between 4 and 5 hours per night, with approximately 50 exposures per sequence.
We note that the first pre-eclipse observation was done during the last commissioning (Program IDs 60.A-9109) in 2023 March 5.
Although this dataset was taken early in the life of NIRPS, its quality is comparable to that of later datasets.

\section{Data processing} \label{sec:data}
NIRPS data are pre-processed by two pipelines: the online NIRPS-DRS running in real time at the telescope, adapted from the ESPRESSO pipeline \citep{pepe_espresso_2021} with telluric correction adapted from \cite{allart_automatic_2022} and A PipelinE to Reduce Observations \citep[APERO;][]{cook_apero_2022} running in Montr\'eal, Canada, originally designed for the SPIRou spectrograph \citep{donati_spirou_2020}.
Both pipelines use similar pre-processing steps including detector effect removal, wavelength calibration, etc. 
They ultimately produce equivalent data products, namely the spectra per order, hereafter called 2D spectra.
One main difference between both pipelines is the telluric-removal process, we refer the reader to Bouchy et al. (submitted) for in-depth analysis of both pipelines.

For our analysis, we use reduced data from the APERO pipeline version 0.7.290. We opted for the 2D telluric-uncorrected spectra and perform the analysis on an order-by-order basis. As we do not use the telluric-corrected spectra, the difference in pipelines should not significantly affect our results. To verify this, we create cross-correlation maps of the detected species using the NIRPS-DRS spectra and obtain the same detections with similar significances (Fig. \ref{fig:apero_vs_geneva}). Furthermore, other analyses using NIRPS have shown little differences in the results when using either pipeline (Bouchy et al. submitted, Allart et al. submitted). We also ran the cross-correlation analysis with the APERO telluric-corrected spectra and we recovered consistent detections, suggesting that our detrending approach is sufficient for telluric line removal.

After pre-processing, we pass the data through the Principal Component Analysis (PCA)-based reduction pipeline described by \citet{pelletier_where_2021} in order to correct telluric and stellar contamination \citep[see also][]{pelletier_vanadium_2023, bazinet_subsolar_2024}.
This method exploits the fact that the spectral lines in the planetary spectrum of WASP-121\,b undergo a considerable shift during the observations due to the change in the line-of-sight velocity of the planet \citep[e.g.][]{birkby_spectroscopic_2018}. 
On the other hand, the velocity shifts of telluric and stellar lines do not change significantly during the same observations. Therefore, the features caused by the Earth's atmosphere or the host star should generally stay on the same pixel on the detector. To keep the planetary signal and remove unwanted contributions, it is thus sufficient to remove features that are constant in wavelength throughout an observation.

\begin{figure}
    \centering
    \includegraphics[width=\linewidth]{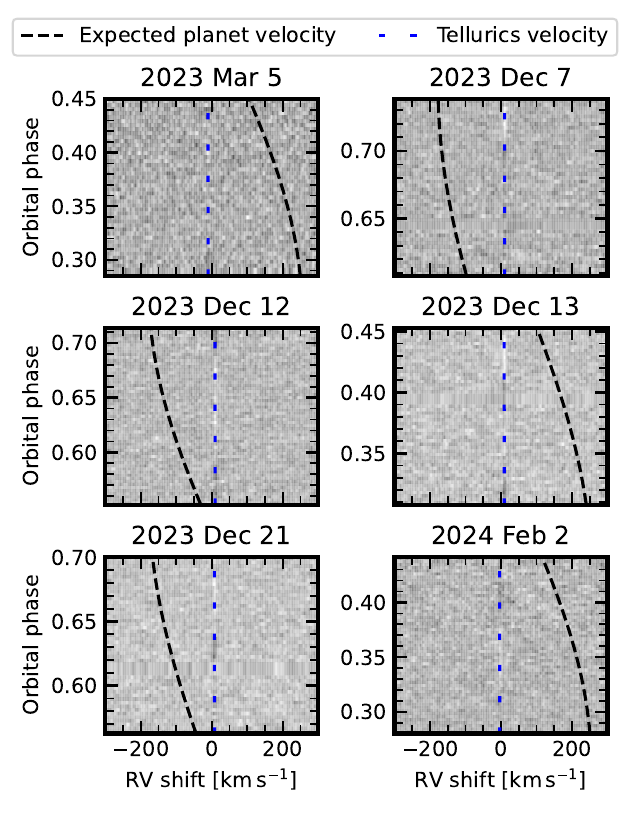}
    \caption{Cross-correlation function of each observation with an OH model (black-and-white colour map), along with the expected WASP-121\,b velocities (black dashed line) compared to where tellurics are in velocity space (blue dotted line). The cross-correlation maps are shown in the BERV-corrected frame.
    In none of our observation nights do the planet velocities overlap with where the telluric are, lowering the risk that any remaining telluric residuals affect our results.
    }
    \label{fig:trail_plots}
\end{figure}

In practice, our approach can be summarised in a handful of steps, applied to each spectral order of each of the six observing sequences.
We start by correcting all pixels that deviate more than 5$\sigma$ from their temporal mean, calculated separately for each observing sequence. 
If three or less flagged pixels are next to each other, we correct them with a linear interpolation using neighbouring pixels. If the group of flagged pixels is bigger than three pixels, they are masked altogether.
We also mask regions with high telluric contamination. The wavelengths between 1.3450 and 1.4465$\,\mu$m are completely masked due to this range being mostly contaminated by atmospheric absorption caused by water vapour. We also mask all lines with depths greater than 70\% of the continuum flux. Major OH emission lines from the Earth's atmosphere are masked using the line list reported by \citet{oliva_lines_2015}. This step still leaves weaker OH lines in the data, which can be numerous in the near infrared. This can affect the detections of OH in the atmosphere of a planet. However, in all six datasets, the expected velocity curve of WASP-121\,b did not fall close to the velocities of the tellurics, such that the spectral lines of WASP-121\,b and of Earth's atmosphere do not overlap (Fig. \ref{fig:trail_plots}). Furthermore, the observations were taken at different times of the year, such that the tellurics do not have the same velocities in the barycentric rest frame between observation.
We continue by discarding orders where the masking has removed more than 80\% of the data.
Afterwards, we apply a high-pass filter to bring each exposure to the same continuum level \citep{gibson_detection_2020}. 
This filter is a two-step process where we first pass a box filter with length 51 pixels, followed by a Gaussian filter with a standard deviation of 100 pixels.
We then divide out a second-order fit of the median spectrum from each exposure.
We use PCA to reconstruct the observed spectral time series using the first five principal components and dividing this out from the data.
This step removes any remaining unwanted telluric and stellar residuals. 
Here we opt to remove five principal components, which is similar to previous works~\citep[e.g.][]{line_solar_2021, mansfield_metallicity_2024, wardenier_phase-resolving_2024, smith_roasting_2024} and is based on what we find best cleans the data of telluric H$_2$O and OH residuals without unnecessarily removing too much of the planetary signal~\citep[e.g.][]{holmberg_first_2022, bazinet_subsolar_2024, pelletier_crires_2024}.
Finally, we mask noisy channels that deviate more than 4$\sigma$ from the standard deviation of their spectral order.

After these steps, the remaining spectra are composed primarily of photon noise, with the planetary signal buried in that intrinsic noise.
The cross-correlation method is then used to stack the signal from many individual spectral lines and recover the unique signature of different chemical species in the planetary atmosphere \citep{snellen_orbital_2010, brogi_signature_2012}.

\section{Modelling and cross-correlation analysis} \label{sec:model}

\begin{figure}
    \centering
    \includegraphics[width=\linewidth]{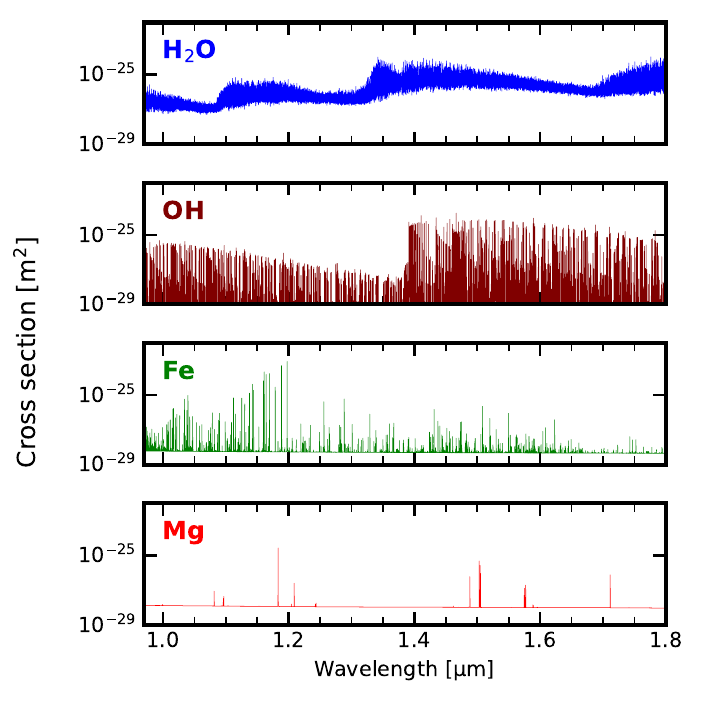}
    \caption{Cross-sections of H$_2$O, OH, Fe and Mg in the wavelength range of NIRPS. The cross-correlations are calculated at a pressure of 1\,mbar and a temperature of 2500\,K.
    H$_2$O contains a forest of lines throughout the entire wavelength range shown here. OH is prominent at wavelengths longer than 1.4\,$\mu$m but still has strong lines around 1\,$\mu$m. Fe features are mostly present in the shorter wavelengths with its most important lines below 1.2$\,\mu$m. Mg has few lines overall.
    }
    \label{fig:crosssections}
\end{figure}

As a first step to reveal the atmospheric composition of WASP-121\,b, we cross-correlate the cleaned NIRPS spectra with generated synthetic atmospheric models.
The atmospheric models used in this analysis were computed using the SCARLET framework \citep{benneke_atmospheric_2012, benneke_how_2013, benneke_strict_2015, benneke_sub-neptune_2019, benneke_water_2019, pelletier_where_2021, pelletier_vanadium_2023, pelletier_crires_2024, bazinet_subsolar_2024}.
Given a vertical temperature profile and a chemical composition, SCARLET can calculate a synthetic line-by-line high-resolution (R = 250,000) emission spectrum of a planet.
One can also let the composition be in chemical equilibrium. SCARLET uses \texttt{FastChem} \citep{stock_fastchem_2018, stock_fastchem_2022} to calculate the mixing ratios of species in this scenario.
In our models, we considered the opacities of H$_2$O \citep{polyansky_exomol_2018}, OH \citep{rothman_hitemp_2010}, Fe, Mg \citep{ryabchikova_major_2015} (Fig. \ref{fig:crosssections}), and H$^{-}$ in both bound-free and free-free form \citep{gray_observation_2021}. SCARLET also considers collision-induced absorption from H$_2$-H$_2$ and H$_2$-He \citep{borysow_collision-induced_2002}.
Initially, other molecular and elemental species were considered in the models. These species include Fe$^{+}$, V, Cr, Mn, Ni, Ca, Co, Sc, Si, Al, Ti, Ti$^{+}$, TiO, CH$_4$, CO, CO$_2$, NH$_3$, HCN, H$_2$S, C$_2$H$_2$, and SiO \citep{mckemmish_exomol_2019,
ryabchikova_major_2015, kurucz_including_2018, kramida_current_2018, rothman_hitemp_2010, li_rovibrational_2015, hargreaves_accurate_2020, yurchenko_exomol_2020, yurchenko_exomol_2022, harris_improved_2006, barber_exomol_2014, chubb_exomol_2020}. 
However, these were never detected in our observations, and they were ultimately not included in the final models.

The model spectra were first convolved with a Gaussian kernel to decrease their resolution to 80,000 to match the NIRPS instrumental resolution. 
Broadening caused by the rotation of the planet is also considered, with the equatorial rotational velocity of WASP-121\,b being $6.99\,$km$\,$s$^{-1}$ \citep[assuming synchronous rotation,][]{bourrier_hot_2020}.
Finally, a box filter is used to account for the planet's orbital motion relative to our line of sight over the course of a single 300\,s or 400\,s exposure, adding a small source of smearing of the planetary lines.

We generate several models with different vertical temperature profiles. These models assume chemical equilibrium with solar elemental composition \citep{asplund_chemical_2021}. The temperatures profiles are composed of four points with pressure levels of $10^{2}$, $10^{-1}$, $10^{-5}$ and $10^{-8}$\,bar. The bottom two points are set to the same temperature, so that the temperature profile between $10^{2}$ and $10^{-1}$\,bar is isothermal. The same is done for the points at $10^{-5}$ and $10^{-8}$\,bar. We vary these two temperatures from 1500 to 5000\,K with a step size of 500\,K, excluding temperature profiles that are entirely isothermal. This ensures that models have a range of temperature gradients, from weak to strong and from inverted to non-inverted.

We cross-correlate each exposure with the planetary models described above at radial velocities that range from $-$400 to 400\,km\,s$^{-1}$ with a step size of 2\,km\,s$^{-1}$.
We phase fold the result of the cross-correlations to produce $K_p-V_{\mathrm{sys}}$ maps, where $K_p$ and $V_{\mathrm{sys}}$ are respectively the planet semi-amplitude velocity and the star's systemic velocity. We consider a grid of Keplerians where the $K_p$ varies from $-$300 to 300\,km\,s$^{-1}$ at a resolution of 1\,km\,s$^{-1}$ and $V_{\mathrm{sys}}$ varies from $-$100 to 100\,km\,s$^{-1}$ at a resolution of 1\,km\,s$^{-1}$.
We sum the cross-correlation coefficients following these Keplerians to create the maps.
To transform the maps to the S/N scaling, we first subtract the map by the mean of the map excluding a $\pm{10}$\,km\,s$^{-1}$ box around the expected velocities to centre the maps around zero. We then divide by the standard deviation of that same region, outside the $\pm{10}$\,km\,s$^{-1}$ box. This is done individually for every map.

\subsection{Cross-correlation results}

\begin{figure*}
    \centering
    \includegraphics[width=\linewidth]{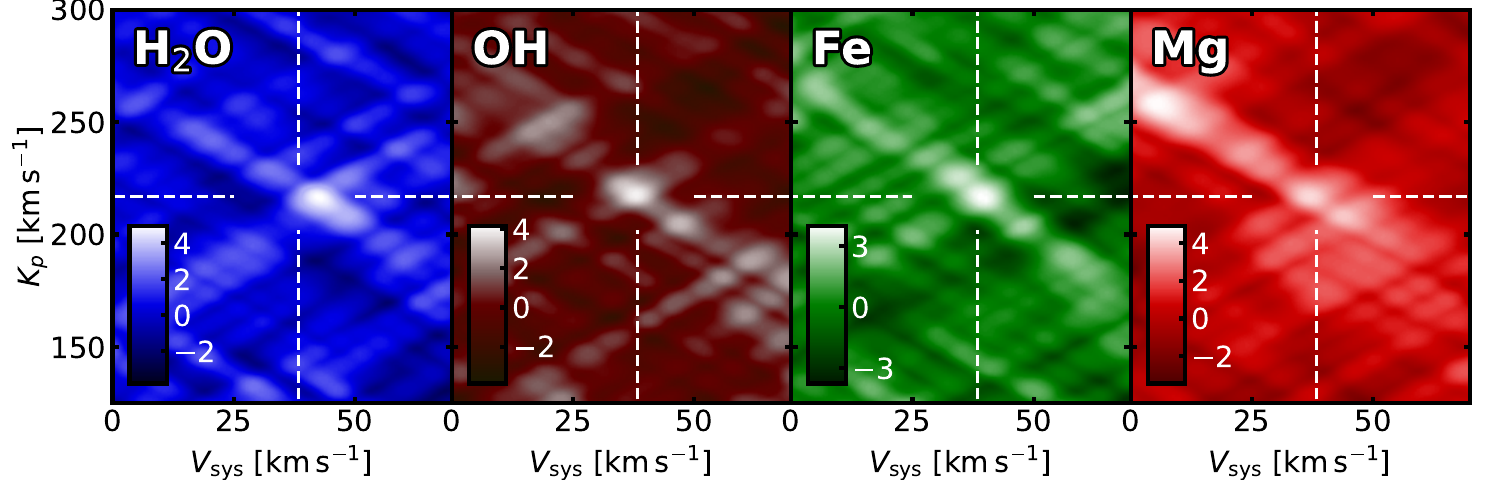}
    \caption{Cross-correlation signal-to-noise maps of H$_2$O, CO, Fe, and Mg in the atmosphere of WASP-121\,b. The dashed vertical and horizontal white lines represent the stellar $V_{\mathrm{sys}}$ \citep[38.12\,km\,s$^{-1}$, from Line-by-line][]{artigau_line-by-line_2022} and $K_p$ \citep[216.8\,km\,s$^{-1}$, calculated from results in][]{bourrier_hot_2020} of WASP-121\,b, respectively.
    Tentative signals of H$_2$O, OH, Fe and Mg can be seen near the expected orbital position of WASP-121\,b.
    }
    \label{fig:kpvsys}
\end{figure*}

Using the cross-correlation method, we find tentative signatures of H$_2$O, OH, Fe and Mg in the dayside atmosphere of WASP-121\,b (Fig. \ref{fig:kpvsys}).
We find that the model that has temperatures of 2000\,K at low altitude and 5000\,K at high altitude produces the strongest detections. 
This model has a temperature inversion, consistent with expectations \citep{evans_ultrahot_2017, changeat_is_2024, hoeijmakers_mantis_2024, pelletier_crires_2024, smith_roasting_2024}.
H$_2$O has a signal with a S/N strength of 4.9, OH with a S/N strength of 3.9, Fe with a S/N strength of 4.0, and Mg with a S/N strength of 4.3. While comparing at the pre-eclipse phases only map and post-eclipse phases map (Fig. \ref{fig:kp_vsys_pre_post}), we see that the pre-eclipse phases have better detections overall.
We note that the strength of the detections does not tell anything about the abundance of the species. We only use the maps as a first step to know which chemical species are present in the NIRPS data. We include the volume mixing ratios of the four species as parameters in the retrieval (Sec. \ref{sec:retrieval}).

While H$_2$O and OH are expected to be present on the dayside of WASP-121\,b and are commonly detected on UHJs with near-infrared spectrographs \citep{nugroho_first_2021, landman_detection_2021, brogi_roasting_2023, gandhi_revealing_2024, mansfield_metallicity_2024, smith_roasting_2024}, the signals observed for Fe are more surprising.
Fe and other metals are typically detected at optical wavelengths, as their opacities are prevalent in these wavelengths.
In this case, however, Fe and Mg lines are observed in the NIRPS bandpass where they still have some spectral lines, detectable with high resolution spectroscopy (Fig. \ref{fig:crosssections}).  While the Fe spectral lines in the near-IR are not as strong as in the optical (Vaulato et al. submitted), the planet-to-star flux contrast in turn is greatly advantageous, vastly increasing the detectability of these species using emission spectroscopy in the near-IR (as opposed to transmission spectroscopy where the area ratio is relatively similar in the near-IR compared to the optical).
Notably, Mg has relatively few lines in the near-IR, but they are preferentially on the redder end of the NIRPS wavelength range compared to Fe where the flux contrast is more favourable.
The cross-correlation map of Mg has a feature at ($K_p$, $\Delta V_{\mathrm{sys}}$) = ($\sim 5$, $\sim 258$) km\,s$^{-1}$. This feature has a higher significance of 4.8 compared to 4.3 for the feature at ($K_p$, $\Delta V_{\mathrm{sys}}$) = ($\sim 37$, $\sim 217$) km\,s$^{-1}$. These two features are aligned on the diagonal trace created by the cross-correlation of the pre-eclipse phases (Fig. \ref{fig:kp_vsys_pre_post}). With few spectral lines, Mg is easily affected by noise and uncertainties, which could explain this abnormality. We cannot exclude that the Mg signal could be an alias of another species, such as Fe.
A recent study found that IGRINS is sensitive to refractories \citep{smith_roasting_2024}. IGRINS covers the $H$ and $K$ bands and therefore does not cover the short infrared wavelengths like NIRPS. This is good evidence that the infrared can be harnessed to probe both the volatiles and refractories in UHJ atmospheres. However, we note that the NIRPS wavelength range does not cover the strong CO bands (at 2.3\,$\mu$m and 4.5\,$\mu$m), making it difficult to detect CO with NIRPS alone. CO is a major molecular constituent in hot and ultra-hot Jupiter atmospheres. As such, a full view of the volatile content of WASP-121b cannot be accurately inferred from these data alone.

We notice $V_{\mathrm{sys}}$ shifts between the signals of some species.
Namely, H$_2$O has a visibly redshifted signal compared to all other species (see Sect. \ref{sec:Vsysshift}).

\section{Retrieval analysis} \label{sec:retrieval}

With the tentative detections of molecular and elemental species in the atmosphere of WASP-121\,b, we proceed with a retrieval to infer the physical properties of this planet.
For this, we use SCARLET coupled with the Bayesian-inference code \texttt{emcee} \citep{foreman-mackey_emcee_2013}.

We use the retrieval to fit the volume mixing ratios of the chemical species found from the cross-correlation analysis. Free constant-with-altitude abundances are fitted for all chemical species.
Since WASP-121\,b is an UHJ, H$_2$O is expected to be severely depleted in the upper atmosphere \citep{parmentier_thermal_2018, pelletier_crires_2024}. As a product of H$_2$O dissociation, OH is also expected to vary considerably as a function of pressure. A constant-with-altitude prescription is therefore not representative of the actual profile of H$_2$O and OH. 
However, the objective of this article is to quantify abundances at photospheric level, so that results obtained at similar pressure levels can be compared.
In the terminator regions of the slightly colder UHJ WASP-76\,b, \citet{gandhi_revealing_2024} found that their OH to H$_2$O ratio did not change when using either the constant-with-altitude prescription or a profile considering dissociation, which supports the idea that a free retrieval is sufficient for our science objective.

We fit the abundance of H$^-$ and e$^-$ as these are expected to be important sources of continuum opacity in UHJ atmospheres \citep[][Vaulato et al. submitted]{arcangeli_h-_2018, parmentier_exoplanet_2018}. These abundances are then used to calculate the H$^{-}$ bound-free and free-free opacity in the models.
We also fit the orbital parameters $K_p$ and $V_{\mathrm{sys}}$.

We introduce a new parameter, $\Delta V_{\mathrm{sys, H_2O}}$, which allows water lines to be Doppler shifted compared to other species in the planet spectrum. This is motivated by the fact that the cross-correlation maps show a redshifted H$_2$O signal, while the other species fall close to the expected velocities (Fig. \ref{fig:kpvsys}).
Indeed, when performing retrievals without the inclusion of this parameter, retrievals would exhibit one of two behaviours. Sometimes, the retrieval would constrain a $V_{\mathrm{sys}}$ near the expected value, with a bounded abundance obtained for OH, but only an upper limit on H$_2$O.
Other times, the retrieval would constrain a $V_{\mathrm{sys}}$ at the redshifted velocity of the H$_2$O signal in the cross-correlation map, with bounded abundance constraints for H$_2$O, but only an upper limit for OH.
In both cases, the velocity separation between H$_2$O and OH prevents the retrieval from well-constraining both species simultaneously.
While it is possible that using a 2- or 3D model that takes into account any potential non-uniform distribution of H$_2$O and OH on the dayside of WASP-121\,b could naturally address this issue, this approach would require prohibitively more computational time and is beyond the scope of this paper.
The $\Delta V_{\mathrm{sys, H_2O}}$ parameter allows both H$_2$O and OH to be simultaneously constrained within our 1D framework, enabling us to better constrain their relative contributions. In particular, relative velocity offsets between different chemical species was observed in UHJ atmospheres~\citep[e.g.][]{brogi_roasting_2023}, which may lead to biases if assuming a single underlying radial shift for all of them.

For the temperature profile, we use the prescription described in \citet{pelletier_where_2021}, where we fit ten temperature points evenly spaced in log pressure between $10^2$ and $10^{-8}$ bar. The smoothing parameter was set to $\sigma_{\mathrm{smooth}}=350\,$K$\,$dex$^{-2}$ and the prior used was uniform between 1 and 8500\,K for each temperature layer.
An overview of the fitted parameters and their priors is shown in Table \ref{tab:res}. 
For the likelihood calculation, as in \cite{pelletier_where_2021}, we inject each generated model (projected in time for a given $K_p$ and $V_\mathrm{sys}$) in a PCA-reconstruction of the data and then passing this through the same detrending steps applied to each spectral order in order to mimic altering effect to the model~\citep{brogi_retrieving_2019, gibson_detection_2020}.

\subsection{Retrieval results}

\begin{center}
\begin{table}
\begin{threeparttable}
\centering
\caption{Retrieved parameters from the free retrieval with their respective priors. $\mathcal{U}(a,b)$ represents a uniform prior from $a$ to $b$.}
\def\arraystretch{1.3}
\begin{tabular}{c c c}
 \hline
 Parameter [unit] & Value retrieved & Prior\\
 \hline
 log$_{10}$H$_2$O & $-3.49^{+0.99}_{-0.93}$ & $\mathcal{U}(-12, 0)$ \\
 log$_{10}$OH & $-3.59^{+1.01}_{-1.08}$ & $\mathcal{U}(-12, 0)$\\
 log$_{10}$Fe & $-2.91^{+1.10}_{-1.36}$ & $\mathcal{U}(-12, 0)$ \\
 log$_{10}$Mg & $-2.75^{+1.09}_{-1.37}$ & $\mathcal{U}(-12, 0)$\\
 log$_{10}$H$^{-}$ & $< -6.06$ (3$\,\sigma$ limit) & $\mathcal{U}(-12, 0)$\\
 log$_{10}$e$^{-}$ & $-7.11^{+4.14}_{-3.40}$ & $\mathcal{U}(-12, 0)$\\
 $K_{p}$ [km\,s$^{-1}$] & $217.21^{+0.63}_{-0.62}$ & $\mathcal{U}(186, 246)$\tnote{1}\\
 $V_{\mathrm{sys}}$ [km\,s$^{-1}$] & $38.04\pm{0.52}$ & $\mathcal{U}(18.36, 58.36)$\tnote{2}\\
 $\Delta V_{\mathrm{sys, H_2O}}$ [km\,s$^{-1}$] & $4.79^{+0.93}_{-0.97}$ & $\mathcal{U}(-15, 15)$\\
 T$_1$,T$_2$, ..., T$_{10}$ [K] & (see Fig. \ref{fig:retrievalres}) & $\mathcal{U}(1, 8500)$\tnote{3} \\
 \hline
\end{tabular}
\begin{tablenotes}
 \item[1] Centred around 216.0\,km\,s$^{-1}$ \citep{hoeijmakers_hot_2020}
 \item[2] Centred around 38.36\,km\,s$^{-1}$ \citep{brown_gaia_2018}
 \item[3] With additional smoothing prior with $\sigma_{\mathrm{smooth}}=350\,$K$\,$dex$^{-2}$ \citep{pelletier_where_2021}
\end{tablenotes}
\label{tab:res}
\end{threeparttable}
\end{table}
\end{center}

\begin{figure*}
    \centering
    \includegraphics[width=\linewidth]{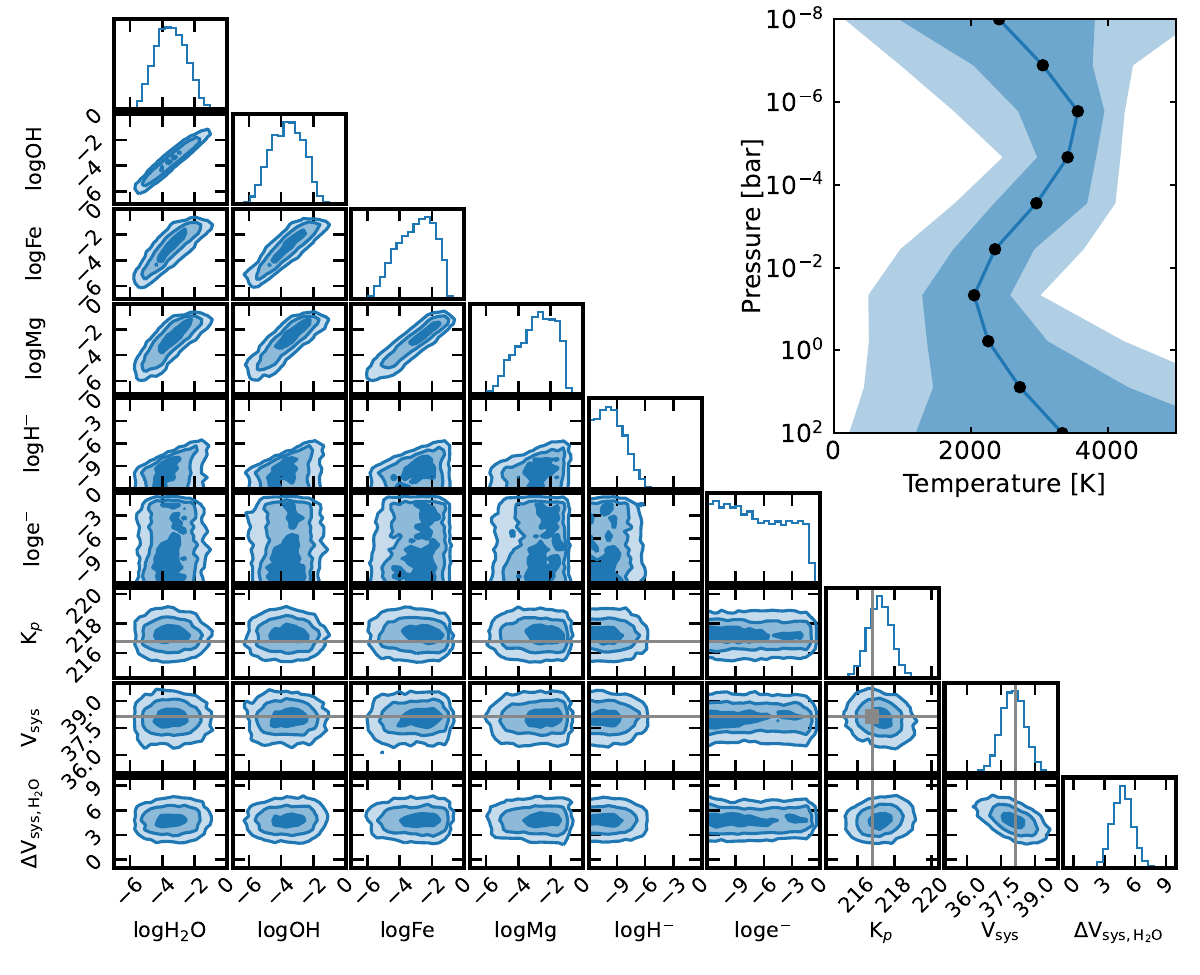}
    \caption{Results from the atmospheric retrieval of WASP-121\,b. The left corner plot shows the posteriors of the retrieved parameters, with the marginalised 1D posteriors on the diagonal and the 2D posteriors on the off-diagonals. The contours on the 2D posteriors represent the 1, 2 and 3\,$\sigma$ uncertainties.
    The first 4 parameters are the base-10 logarithm of volume mixing ratios of the four detected species: H$_2$O, OH, Fe and Mg. The 1D posteriors for the retrieved abundances show relatively broad distributions. However, 2D posteriors between abundances depict a strong positive correlation, indicating that the abundance ratios are better constrained.
    logH$^{-}$ and loge$^{-}$ are the volume mixing ratio of the H$^{-}$ radical and of electrons, respectively, used as broad band opacities. 
    Only a weak upper bound is obtained on logH$^{-}$ while the posterior abundance of loge$^{-}$ spans the full prior.
    The grey lines on the $K_p$ and $V_{\mathrm{sys}}$ parameters are the expected orbital velocities of the planet \citep{bourrier_hot_2020, brown_gaia_2018}. Both orbital parameters are consistent with expectations.
    $\Delta V_{\mathrm{sys, H_2O}}$ is the additional $V_{\mathrm{sys}}$ shift added to H$_2$O. The posterior is constrained to a value above 0, indicating that the retrieval also finds the shift seen in the cross-correlation maps (Fig. \ref{fig:kpvsys}).
    The top right plot is the retrieved vertical temperature profile, with the shaded areas representing the 1 and 2\,$\sigma$ uncertainties. The profile has a temperature inversion, which is expected for UHJs.}
    \label{fig:retrievalres}
\end{figure*}

Using a free atmospheric retrieval, we constrain the abundances of H$_2$O, OH, Fe and Mg, as well as the velocities of each species (Fig. \ref{fig:retrievalres}). The retrieved volume mixing ratios are log$_{10}$H$_2$O $= -3.49^{+0.99}_{-0.93}$, log$_{10}$OH$= -3.59^{+1.01}_{-1.08}$, log$_{10}$Fe$= -2.91^{+1.10}_{-1.36}$, and log$_{10}$Mg$= -2.75^{+1.09}_{-1.37}$.
The uncertainties on the retrieved absolute abundances are relatively large due to the strong correlation between abundance and the temperature-pressure profile. However, relative abundances largely unaffected by this degeneracy and are therefore less dependent on model assumptions \citep[e.g.][]{brogi_retrieving_2019, line_solar_2021, gibson_relative_2022, gandhi_retrieval_2023, maguire_high-resolution_2023}.
We retrieved a volume mixing ratio of log$_{10}$H$^{-} < -6.06$ (3$\,\sigma$ limit) while log$_{10}$e$^{-}$ is not well constrained, with the posterior spanning the entire prior space. 

The Doppler shifts of all species are retrieved with a $K_p$ value of $217.21^{+0.63}_{-0.62}\,$km$\,$s$^{-1}$, which is consistent with the known semi-amplitude velocity of WASP-121\,b considering the planet's rotation \citep[$216.8 \pm 4.5$\,km\,s$^{-1}$,][]{bourrier_hot_2020}. We will further comment on this result in Sect. \ref{sec:Kpshift}.

The retrieval found a systemic velocity for the combined OH, Fe and Mg signal of $38.04\pm{0.53}\,$km$\,$s$^{-1}$,
consistent with the systemic velocity retrieved by the Line-by-line technique using the NIRPS observations presented in this paper as well as three transmission dataset~\citep[38.12\,km\,s$^{-1}$,][]{artigau_line-by-line_2022}.
Our retrieved value is also consistent with previous WASP-121\,b results \citep{brown_gaia_2018, bourrier_hot_2020}.
We quantify the shift of the H$_2$O velocity distribution using the retrieval and measure a constrained value of $4.79^{+0.93}_{-0.97}\,$km$\,$s$^{-1}$. The retrieval therefore finds the $V_{\mathrm{sys}}$ of the H$_2$O signal to be at $42.83^{+0.80}_{-0.79}\,$km$\,$s$^{-1}$.
We attribute this anomalous shift in the water lines to the 3D nature of the atmosphere of WASP-121\,b. We will discuss potential mechanisms causing this shift in Sect. \ref{sec:Vsysshift}.

We retrieve a vertical temperature profile with an inversion layer, as predicted from the cross-correlation detections. This inversion is consistent with modelling and dayside observations of WASP-121\,b \citep{evans_ultrahot_2017, changeat_is_2024, hoeijmakers_mantis_2024, pelletier_crires_2024, smith_roasting_2024}.\\

\section{Discussion} \label{sec:discussion}

\subsection{Water dissociation}

At the extreme temperatures prevailing on the daysides of UHJs, some of the atmospheric molecular constituents are expected to thermally dissociate. This is the case for H$_2$O, which dissociates into elemental H and O \citep{parmentier_thermal_2018}. These elements recombine to produce OH which, when present in an appreciable amount in the atmosphere, is detectable with high-resolution spectroscopy \citep{nugroho_first_2021, landman_detection_2021, finnerty_keck_2023, brogi_roasting_2023, mansfield_metallicity_2024, smith_roasting_2024}.
The reaction chain is as follows:\\

\noindent H$_2$O $\rightleftarrows$ 2H + O\\
H + H + M $\rightarrow$ H$_2$ + M\\
O + H$_2$ $\rightarrow$ H + OH, \\

\noindent
where M is any chemical species \citep{parmentier_thermal_2018}. OH is involved in the recombination of molecular species such as TiO. However, these recombination reactions are significantly slower than the recombination of OH and are not expected to remove a substantial amount of OH in the atmosphere of UHJs \citep{parmentier_thermal_2018}.

Another source of OH in an atmosphere is the photolysis of H$_2$O by UV photons from the host star \citep{liang_source_2003}.
Disequilibrium processes such as photodissociation are believed to be more prominent on colder hot Jupiters and become less important as temperature increases, with UHJ atmospheres being well approximated by thermochemical processes alone
\citep{roudier_disequilibrium_2021, moses_disequilibrium_2011}. 
A recent study by \citet{baeyens_photodissociation_2024} looked at the photodissociation of numerous molecular species, including H$_2$O, in the atmosphere of the UHJ WASP-76\,b. They concluded that the production of OH caused by the photolysis of H$_2$O becomes significant at pressures lower than $10^{-5}\,$bars (0.01\,mbar).
Given that emission and transmission spectroscopy usually probes between the $10^0$ to $10^{-3}$\,bars pressure levels \citep[e.g.][]{smith_roasting_2024, gandhi_revealing_2024}, which is significantly higher pressures than the 10$^{-5}$\,bars balance point, it is safe to assume that the OH and H$_2$O probed in our observations are mostly affected by thermal dissociation rather than photodissociation.
Theoretical work on other hot Jupiters has found similar results, where disequilibrium chemistry is only significant at the top of the atmosphere \citep[e.g.][]{molaverdikhani_cold_2019, moses_disequilibrium_2011, moses_chemical_2014}.
WASP-121\,b orbits a star similar to that of WASP-76\,b, however it has a shorter orbital period. 
As such, its equilibrium temperature (T$_\mathrm{eq} \sim 2350$\,K) is about 150\,K hotter than WASP-76\,b (T$_\mathrm{eq} \sim 2200$\,K). We should thus expect more thermal dissociation on WASP-121\,b compared to WASP-76\,b, if probing at the same pressure level. This would make it even more difficult for OH from photodissociation to overcome its thermally dissociated counterpart.

\begin{figure*}
    \centering
    \includegraphics[width=0.85\linewidth]{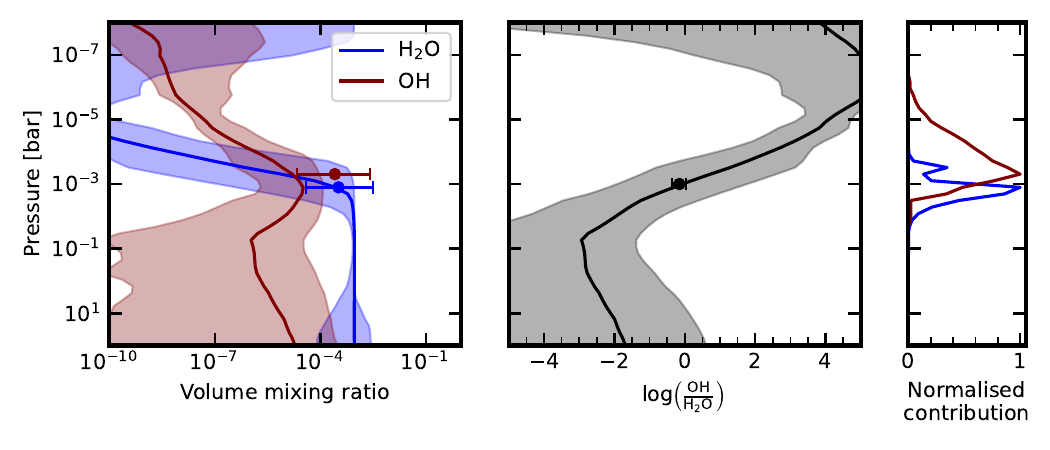}
    \caption{Comparison between retrieved molecular abundances and equilibrium predictions. We create models in chemical equilibrium using \texttt{FastChem} \citep{stock_fastchem_2022, stock_fastchem_2018} with a 5 times solar metallicity composition and where the vertical temperature profile is sampled from the retrieval.
    \textit{Left:} The curves and the shaded areas represent the median and 1\,$\sigma$ uncertainties of the mixing ratios of H$_2$O and OH from the equilibrium models. The points are the retrieved abundance for H$_2$O and OH, with their y height at the approximate pressure that they probe.
    \textit{Middle:} The line and the shaded regions are the median and 1\,$\sigma$ uncertainties on the (base-10) log ratio of OH/H$_2$O calculated from the equilibrium models. The point represent the $\log_{10}(\mathrm{OH}/\mathrm{H_2O})$ calculated from the retrieval along with the associated 1\,$\sigma$ uncertainties. The point is at 1\,mbar, which is the approximate pressure level where the observations probe.
    \textit{Right:} Contribution curves depicting where in the atmosphere the signals of H$_2$O (blue) and OH (maroon) are probed. Curves are normalised so that their maximum is equal to one.
    }
    \label{fig:retrieval_vs_chemequi}
\end{figure*}

We attempt to quantify the dissociation of H$_2$O by calculating the ratio of abundance of OH and H$_2$O. However, we note that while this approach may seem intuitive, there are some caveats to consider. First, the abundances of H$_2$O and OH in the atmosphere of WASP-121b are expected to decrease at sub-millibar pressures (Figure~\ref{fig:retrieval_vs_chemequi}, left panel).  Such a gradient in chemical abundance would result in a decrease in line strengths~\citep{parmentier_thermal_2018}, which our constant-with-altitude abundance profile models cannot reproduce and may fit for lower abundances to compensate. Secondly, the average pressures probed by H$_2$O and OH lines are not identical (Figure~\ref{fig:retrieval_vs_chemequi}, right panel), which means that their ratio may not accurately represent a given pressure level.
Nevertheless, with these caveats in mind, our retrieval shows a ratio of log$_{10}$(OH/H$_2$O) $= -0.15\pm{0.20}$. If we assume that there is a one-to-one relationship between the dissociated H$_2$O and the OH production, this result indicates that $41\pm{11}\%$ of the H$_2$O is dissociated\footnote{The dissociation fraction is calculated as $(10^{-\log_{10}(\mathrm{OH}/\mathrm{H_2O})} + 1)^{-1}$.}.

We start by comparing the retrieved abundances to prediction from chemical equilibrium models. We generate several chemical equilibrium models by assuming a temperature profile sampled from the retrievals. Models have a 5 times solar metallicity, following the results from \citet{smith_roasting_2024}, who calculated the metallicity using molecular species such as H$_2$O and OH. With these models, we can find the mean and uncertainties on the abundance of H$_2$O and OH at every pressure level (Fig. \ref{fig:retrieval_vs_chemequi}). We further calculate the contribution functions for an atmosphere corresponding to the retrieval median fit. These contribution functions show that both H$_2$O and OH probe close to the mbar level (Fig. \ref{fig:retrieval_vs_chemequi}). The retrieved mixing ratio of H$_2$O is highly consistent with the chemical equilibrium prediction at 1\,mbar. However, the H$_2$O abundance decreases rapidly at pressures just below the mbar level. The abundance at 1\,mbar is around four orders of magnitude greater than at 0.1\,mbar.
The retrieved OH abundance is consistent within 1\,$\sigma$ with the prediction from chemical equilibrium at 1\,mbar.
From these chemical models, we calculate the predicted OH/H$_2$O ratio at each pressure layer (Fig. \ref{fig:retrieval_vs_chemequi}). This ratio varies considerably in the mid-atmosphere, where log$_{10}$(OH/H$_2$O) $\sim -2$ at 100\,mbar and log$_{10}$(OH/H$_2$O) $\sim 4$ at 0.01\,mbar. The predicted log ratio at 1\,mbar is $\sim 0$. This is highly consistent with the retrieved ratio of $-0.15\pm{0.20}$. This shows that chemical equilibrium models are good predictors of thermal water dissociation, in agreement with modelling work that predicts UHJs should have atmospheres dominated by thermochemical processes \citep{roudier_disequilibrium_2021, moses_disequilibrium_2011}.

However, the above discussions assume that WASP-121\,b's dayside composition is homogeneous throughout the dayside, which might not be realistic. The cross-correlation map shows different dynamics for OH and H$_2$O.
We could therefore be retrieving the H$_2$O in a region of the atmosphere where there is, for example, a higher abundance than at the location where OH is retrieved. This could also be said for the altitude. However, the contribution functions show that they mostly probe the same pressure levels, with OH probing only slightly lower pressures (Fig. \ref{fig:retrieval_vs_chemequi}).

\begin{figure}
    \centering
    \includegraphics[width=\linewidth]{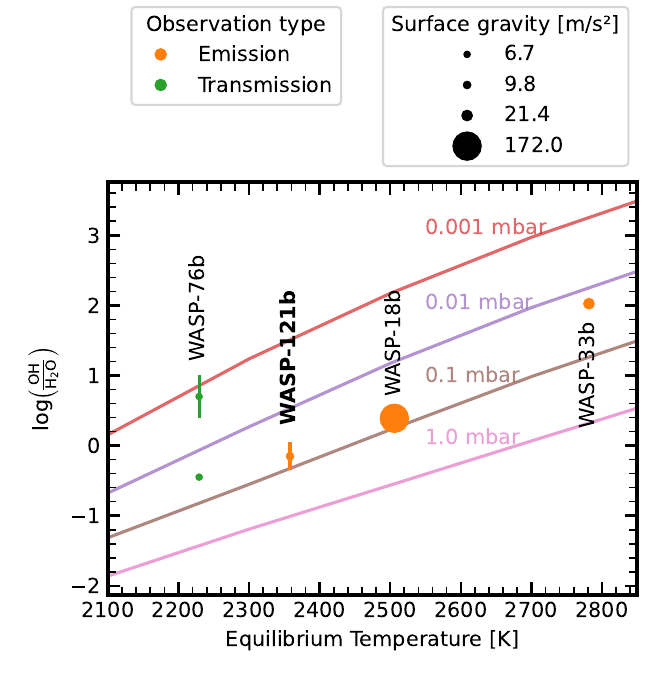}
    \caption{The ratio of the abundance of OH over the abundance of H$_2$O as a function of equilibrium temperature from high-resolution retrievals. The orange points are from the dayside emission of the planets WASP-121\,b (this work), WASP-18\,b \citep[left orange point,][]{brogi_roasting_2023}, and WASP-33\,b \citep[right orange point,][]{finnerty_keck_2023}. The green points are obtained from the transit of WASP-76\,b \citep{mansfield_metallicity_2024, gandhi_revealing_2024}. Points with error bars are from papers where the value and uncertainty on log$_{10}$(OH/H$_2$O) are reported \citep[][this work]{gandhi_revealing_2024}. For points without error bars, we take the median of the posterior of OH and H$_2$O to calculate the ratio.
    The area of the points is proportional to the surface gravity of the planets.
    The lines are the prediction from chemical equilibrium models computed using \texttt{FastChem} \citep{stock_fastchem_2022, stock_fastchem_2018} at different pressure levels. The models used are atmospheric models of WASP-121\,b, where we set the metallicity and C/O to a solar value and with a uniform vertical temperature profile at the corresponding equilibrium temperature.
    }
    \label{fig:OH_over_H2O}
\end{figure}

\citet{gandhi_revealing_2024} retrieved the ratio of OH over H$_2$O in the terminator of the colder UHJ WASP-76\,b using combined CARMENES and Hubble WFC3\footnote{Wide Field Camera 3} data. Contrary to our results on WASP-121\,b, they found that the H$_2$O is mostly dissociated with an abundance ratio of $\log_{10}($OH/H$_2$O$) = 0.7\pm{0.3}$.
While it is true that thermal dissociation should be overall more prevalent on hotter planets, the pressure levels probed is also an important factor (Fig. \ref{fig:OH_over_H2O}). Dissociation is more prominent at lower pressures (or higher altitudes). Since transmission spectroscopy generally probes lower pressures than emission spectroscopy, it is not necessarily surprising that dissociation may appear more pronounced relative to emission spectroscopy. 
However, the transmission data analysed by \citet{gandhi_revealing_2024} probe an average pressure level of 1.5\,mbar while the OH detected from our dayside observations originates from around 1\,mbar and extends at even lower pressures, contradicting the usual assumption that transmission probes at higher altitude. This discrepancy could be caused by a difference in surface gravity or other such planetary parameter between both planets.
\citet{mansfield_metallicity_2024} were also able to find the abundance of H$_2$O and OH in the terminator of WASP-76\,b using IGRINS data. They found that the H$_2$O is dominant by about half to one order of magnitude, depending on the vertical abundance parameterisation used.
While this appears to be inconsistent with the results of \citet{gandhi_revealing_2024}, it is unclear whether the discrepancy between these works may simply be due to their different modelling approaches.

We also compare our results with other UHJ daysides where H$_2$O and OH were found simultaneously using high resolution data (Fig. \ref{fig:OH_over_H2O}). The hottest UHJ on which OH has been detected on its dayside is WASP-33\,b \citep{nugroho_first_2021}. There, OH was found to be two orders of magnitude more abundant than H$_2$O \citep{finnerty_keck_2023}. 
On the colder UHJ WASP-18\,b, \citet{brogi_roasting_2023} retrieved a constrained H$_2$O abundance but a lower limit for the OH abundance. They exclude (at 2$\,\sigma$) scenarios where H$_2$O is three orders of magnitude more abundant than OH. 
Their median fit implies that $\log_{10}($OH/H$_2$O$) \sim 0.4$. 
The coldest UHJ where the dayside OH and H$_2$O abundances were retrieved is WASP-121\,b, with the results shown in this work.
Taken together, these five data points show a possible trend, indicating that warmer planets have more OH, and therefore more thermal dissociation. This is logical since dissociation is directly linked to temperature \citep{parmentier_thermal_2018}.
However, we note that this trend is expected to fall off at extremely high temperatures. At a certain point, OH is expected to dissociate as well, forming elemental O and H \citep[e.g.][]{landman_detection_2021, brogi_roasting_2023, gandhi_revealing_2024}.
The planets follow closely the trend lines from chemical equilibrium models (Fig. \ref{fig:OH_over_H2O}). The observations follow the trends from equilibrium models at pressures between 0.1\,mbar and 0.01\,mbar. This does not mean that this is the pressure that is probed by emission spectroscopy. These models assume a uniform temperature profile equal to the equilibrium temperature. On a hot Jupiter, the dayside temperature is higher than the equilibrium temperature. Therefore, the models do not reflect the actual temperature profile in the atmosphere of the planets.
Nevertheless, the trend lines can still be useful to compare with observations from a qualitative point of view.
Another factor affecting the observed thermal dissociation is the surface gravity of the planet \citep{parmentier_thermal_2018}. A difference in gravity will change the pressure level of the photosphere. For a low gravity planet such as WASP-121\,b, the photosphere is expected to be at lower pressure, where thermal dissociation of H$_2$O is more prominent, although this is not what is portrayed in Fig. \ref{fig:OH_over_H2O}. WASP-121\,b has lower gravity than WASP-33\,b and considerably lower gravity than WASP-18\,b. Nevertheless, WASP-33\,b, WASP-18\,b and WASP-121\,b all seem to trace a similar trend line, indicating that the surface gravity might not be such an important factor when it comes to the observations of dissociation in UHJ daysides. However, as all of these measurements were obtained from independent analyses of data sets taken with different instruments, we caution that this mask underlying trends in the observations. This is evidenced by the seemingly different OH/H$_2$O ratio inferred for WASP-76b~\citep{mansfield_metallicity_2024, gandhi_revealing_2024}. Further uniform analyses with surveys from the same instrument reduced by the same pipeline may shed more insights into the degree of thermal dissociation in different ultra-hot Jupiter atmospheric conditions.

Overall, OH is still a relatively newly discovered molecule in the atmospheres of UHJs, with only a handful of planets in which its abundance relative to H$_2$O has successfully been retrieved \citep{finnerty_keck_2023, brogi_roasting_2023, mansfield_metallicity_2024, gandhi_revealing_2024}. There is still much to learn regarding the trends seen in the thermal dissociation of molecules in UHJ atmospheres. As more OH is detected in other planets with high resolution infrared spectrograph such as NIRPS, a clearer picture of how dissociation plays a role in shaping the population of UHJ atmospheres will emerge.

\subsection{Velocity traces} \label{sec:shifts}

The $K_p-V_{\mathrm{sys}}$ maps show signals that are not at the expected planetary velocities (Fig. \ref{fig:kpvsys}).
Firstly, the retrieved $K_p$ for all species is smaller than the expected planetary $K_p$.
Secondly, the H$_2$O signal is noticeably redshifted compared to other detected species. 
In this subsection, we investigate these shifts using Global Circulation Models (GCMs). We conclude that the $K_p$ shift is conceptually predicted by the planet's rotation, meanwhile, the $V_{\mathrm{sys}}$ relative shift cannot be fully explained by GCMs.

\subsubsection{Global Circulation Models} \label{sec:GCM}

To further interpret the observed Doppler shifts, we compare our results to the Doppler shifts predicted by four 3D Global Circulation Models (GCMs) of WASP-121\,b. The first GCM is the drag-free model presented in \citet{parmentier_thermal_2018}, while the other three models are based on recent work by \citet{tan_modelling_2024}. The latter includes a drag-free model, a model with weak drag ($\tau_{\mathrm{drag}}=10^6$\,s) and a model with strong drag ($\tau_{\mathrm{drag}}=10^4$\,s), where $\tau_{\mathrm{drag}}$ is the uniform drag timescale of the atmosphere. The models from \citet{tan_modelling_2024} account for additional heat transport between the dayside and the nightside due to hydrogen dissociation/recombination (see also \citealt{bell_increased_2018, tan_atmospheric_2019, roth_pseudo-2d_2021}), which impacts the 3D temperature structure of the planet. For a more in-depth discussion of the GCMs we refer to \citet{wardenier_phase-resolving_2024}, who used the same models to interpret transmission observations of WASP-121\,b performed with IGRINS.

To obtain phase-dependent emission spectra for each of the four GCMs, we post-process the models with gCMCRT \citep{lee_mantis_2022, lee_3d_2022}, a 3D Monte-Carlo radiative-transfer code. gCMCRT is able to account for the Doppler shifts imparted on the opacities in each atmospheric cell by considering the line-of-sight velocities due to the planet's wind profile and its rotation. For details regarding the inner workings of the code, we refer to \citet{lee_3d_2022} and \citet{wardenier_pretransit_2025}. We calculate the emission spectra for each of the models during pre-eclipse (phases 0.29 to 0.46) and post-eclipse (phases 0.54 to 0.71), with intervals of 10 degrees in orbital phase, at a spectral resolution of 135,000. The radiative transfer is performed in the IGRINS bandpass and considers the same line/continuum species as those used in \citet{wardenier_modelling_2023,wardenier_phase-resolving_2024}. This includes H$_2$O and OH, which are relevant for our study. Although IGRINS (1.42 - 2.42\,$\mu$m) and NIRPS (0.98 - 1.80\,$\mu$m) have different bandpasses, we expect our models to be equally representative for NIRPS. This is because both bandpasses cover a dense forest of H$_2$O lines which should, on average, probe similar pressures. Furthermore, the cross-sections of OH are strongest in the wavelength region beyond 1.4\,$\mu$m, where the instruments overlap (see Fig. \ref{fig:crosssections}).

\begin{figure*}
    \centering
    \includegraphics[width=\linewidth]{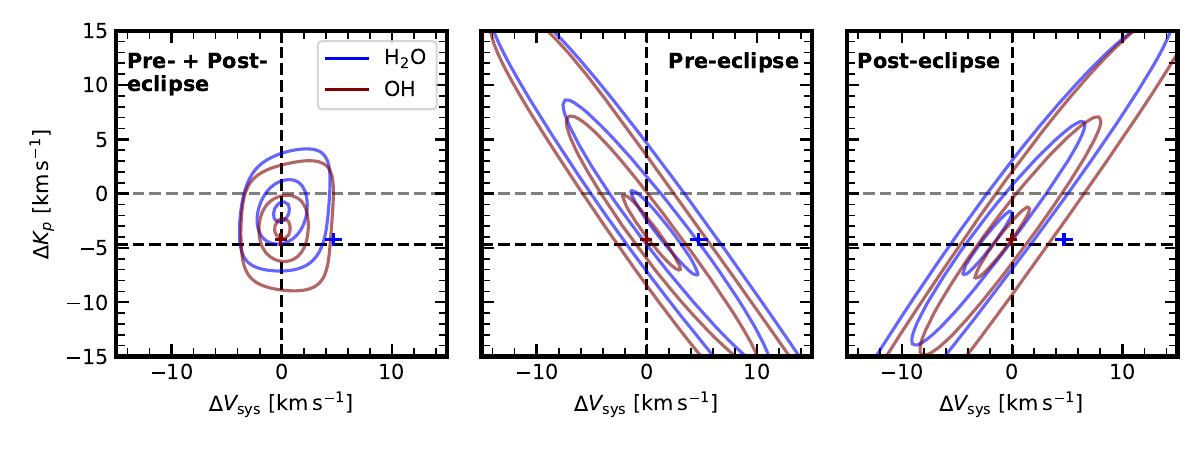}
    \caption{Contour plot of predicted $K_p-V_{\mathrm{sys}}$ maps by a WASP-121\,b GCM. The model used is a strong drag model ($\tau_{\mathrm{drag}}=10^4$\,s) based on the work of \citet{tan_modelling_2024}. 
    The black dashed lines at ($\Delta K_p$, $\Delta V_{\mathrm{sys}}$) = (0, $\sim -5$) km\,s$^{-1}$ represent the planet position considering the rotation of the planet and assuming that the observed signal is uniform throughout the visible hemisphere.
    The grey line denotes the expected $K_p$ without the rotation of the planet.
    The blue and maroon crosses are the retrieved velocities of H$_2$O and OH, respectively, along with their 1\,$\sigma$ uncertainties. To get the contour lines, we first cross-correlate noiseless GCMs with molecular opacity templates, followed by a phase folding in the $K_p-V_{\mathrm{sys}}$ space. The contours represent an elevation of 0.99, 0.9 and 0.7 times the map maximum.
    The left panel is the resulting map from taking both the pre- and post-eclipse phases. The middle and right plot consider only the pre-eclipse and post-eclipse, respectively.
    }
    \label{fig:GCM_strongdrag}
\end{figure*}

Once all emission spectra are obtained, we compute the $K_p-V_{\mathrm{sys}}$ maps associated with H$_2$O and OH in pre-/post-eclipse by cross-correlating the spectra with a template spectrum and co-adding the resulting CCFs along the phase axis (similar to \citealt{wardenier_phase-resolving_2024}). The results are shown in Fig. \ref{fig:GCM_strongdrag}. Because \mbox{gCMCRT} ignores the orbital motion of the planet when computing the spectra, the Doppler shifts shown in the plots are inherently due to the 3D structure and dynamics of the atmosphere.

The GCMs predict a decrease in the observed $K_p$, regardless of the model used. See Sect. \ref{sec:Kpshift} for a detailed discussion of this effect.
On the other hand, GCMs are not able to reproduce the magnitude of the $V_{\mathrm{sys}}$ redshift observed in the case of H$_2$O.
However, certain scenarios, such as considering only the pre-eclipse phases, can lead to a small $V_{\mathrm{sys}}$ shift between H$_2$O and OH. Possible explanations are discussed in Sect. \ref{sec:Vsysshift}.

\subsubsection{Planetary semi-amplitude $K_p$} \label{sec:Kpshift}

We retrieve a planetary semi-amplitude which is consistent with expectations. The retrieved $K_p$ of $217.21^{+0.63}_{-0.62}\,$km$\,$s$^{-1}$ is well consistent with $216.8\pm 4.5\,$km$\,$s$^{-1}$ \citep{bourrier_hot_2020}, which takes into consideration the effect of the rotation of WASP-121\,b. To calculate the expected $K_p$, we use the following equation, which assumes a circular orbit:

\begin{equation} \label{eq:Kp_w_rot}
     K_p = \left(V_{\mathrm{orb}} - \frac{2}{3} V_{\mathrm{rot}}\right) \sin i = \frac{2\pi}{P} \left(a - \frac{2}{3}R_p\right) \sin i
\end{equation}

\noindent where $V_{\mathrm{orb}} = 2\pi a/P$ is the linear velocity of the planet in its orbit and $V_{\mathrm{rot}} = 2\pi R_p/P$ is the equatorial rotational velocity assuming synchronous rotation. 
Traditionally, the rotational velocity, which counters the orbital velocity of the planet, is not considered in the calculation of $K_p$. However, for large planets on small orbits, the effect of the rotational velocity on dayside observations can be in the order of a percent \citep[][see their Fig.\ 8]{hoeijmakers_mantis_2024}. 

The effect of the rotation of a large planet on its radial-velocity semi-amplitude can be conceptually modelled.
The equation dictating the radial velocity of a point of the equator at longitude $\theta$ (in radians) on a surface of a zero-eccentricity planet at phase $\phi$ (in radians) is given by \citep{hoeijmakers_mantis_2024}:

\begin{equation} \label{eq:rv_on_pla}
    v_r(\phi, \theta) = V_{\mathrm{orb}} \sin(\phi) - V_{\mathrm{rot}} \sin(\phi - \theta).
\end{equation}

\noindent This equation also supposes that the planet is tidally-locked and has an inclination $i \sim 90^{\circ}$\footnote{If $i$ is not close to $90^{\circ}$, all equations must be multiplied by $\sin i$.}.
When not considering the rotation (i.e. $V_{\mathrm{rot}} = 0$), we get that $K_p =  V_{\mathrm{orb}}$, which is what is mostly used in literature.
However, when considering a rotation on the planet, the expected semi-amplitude will vary.
Setting $\theta = 0$, we see that the measured $K_p$ should be $V_{\mathrm{orb}} - V_{\mathrm{rot}}$.
One can visualise this decrease in velocity by imagining the velocity of the substellar point throughout the orbit. This point will always be in the interior of the orbit. Therefore, it draws a smaller orbit around the star, hence it will have a smaller $K_p$ compared to the centre of mass of the planet.
However, the signal is never emitted from only one point in the atmosphere.
We rather get the signal as a weighted average of the visible hemisphere of the planet. 
If we suppose that the signal was coming equally throughout the dayside, we expect a decrease of $\frac{2}{3}V_{\mathrm{rot}}$ when observing the planet close to the eclipse. This assumption is made in the calculation of the expected planetary semi-amplitude and is the origin of the $\frac{2}{3}$ factor in Eq. \ref{eq:Kp_w_rot}.
However, the weighting depends not only on the flux emitted but also the line depth at each point in the visible atmosphere. As high-resolution spectroscopy probes the depth of chemical lines as opposed to the continuum level, it is therefore important to understand the relationship between the vertical temperature profile and the resulting spectra. Although counterintuitive, the region where the temperature is the highest, the hot spot, is not always where most of the signal is coming from. 
It was shown that an UHJ with an eastern hot spot offset has an eastern dayside that is usually more isothermal compared to the western dayside \citep{van_sluijs_carbon_2023}.
This means that, even with a colder temperature, the difference between the line and continuum in the western dayside is greater, and the signal is therefore coming mainly from that region \citep{van_sluijs_carbon_2023}. 
We are not suggesting that this is the case for WASP-121\,b, just that the measured $K_p$ is not trivial to calculate, since it requires the complete temperature profile of the visible hemisphere for the entire duration of the observation.

Since GCMs consider Doppler shifts induced by the rotational velocity of WASP-121\,b, they predict the $K_p$ decrease caused by rotation regardless of the treatment of winds (Appendix Fig. \ref{fig:GCMs_4_models}). This effect is also seen for pre-eclipse and post-eclipse phases individually (Fig. \ref{fig:GCM_strongdrag}).

Our retrieved planetary semi-amplitude, $217.21^{+0.63}_{-0.62}\,$km$\,$s$^{-1}$, is consistent within 2\,$\sigma$ with \citet{hoeijmakers_mantis_2024}, where they found a value of $\sim$216\,km\,s$^{-1}$ using optical observations of the dayside of WASP-121\,b. Our $K_p$ is also consistent with other emission analysis of WASP-121\,b \citep{smith_roasting_2024, pelletier_crires_2024}.

A caveat of the discussion above is the uncertainty on the value of the planet semi-amplitude velocity, $K_p$.
Without considering the velocity from a the planet's rotation, one can calculate $K_p$ using one of two equations \citep{birkby_spectroscopic_2018}:
\begin{equation} \label{eq:Kp_vorb}
    K_p = V_{\mathrm{orb}} \sin i = \frac{2 \pi a}{P} \sin i
\end{equation}
\noindent or
\begin{equation} \label{eq:Kp_K}
    K_p = \frac{M_*}{M_p}K_*.
\end{equation}
\noindent These two equations should give the same value for $K_p$. However, when using the median values reported by \citet{bourrier_hot_2020}, Eq. \ref{eq:Kp_vorb} gives $K_p = 221.4 \pm 4.5$\,km$\,$s$^{-1}$ \citep[used in e.g.][]{pelletier_crires_2024, hoeijmakers_mantis_2024}, while Eq. \ref{eq:Kp_K} gives $K_p = 218^{+17}_{-16}$\,km$\,$s$^{-1}$ \citep[used in e.g.][]{wardenier_phase-resolving_2024, smith_roasting_2024}. 
Although these values are consistent, there is a major difference between the uncertainties. This difference stems from the uncertainties of the underlying observables. Equation \ref{eq:Kp_vorb}'s uncertainty is dominated by the semi-major axis $a$. In \citet{bourrier_hot_2020}, $a$ is calculated using $(a/R_*)R_*$. Therefore, the underlying uncertainty comes from the value of the stellar radius, which is $\sim 2\%$.
As for Eq. \ref{eq:Kp_K}, the uncertainty on $K_p$ comes from the uncertainties in the radial velocity measurements. $M_*$, $M_p$ and $K_*$ have uncertainties of $\sim 5\%$, which leads to an uncertainty on $K_p$ of $\sim 7.5\%$, which is more than three times larger than using Eq. \ref{eq:Kp_vorb}.
In literature, the uncertainty on $K_p$ is rarely considered. However, it is crucial to take it into consideration, since a shift in $K_p$ of a few km\,s$^{-1}$ may be incorrectly explained by atmospheric dynamics even if it is still consistent with the expected value.

\subsubsection{$V_{\mathrm{sys}}$ shifts} \label{sec:Vsysshift}

The visible redshift of the H$_2$O signal is detected with a value of $4.79^{+0.93}_{-0.97}\,$km$\,$s$^{-1}$, which is $\sim 5\,\sigma$ away from zero. 
We also observe small $V_{\mathrm{sys}}$ shifts in the Fe and Mg signals ($\sim \pm 1\,$km$\,$s$^{-1}$). However, we focus our efforts on H$_2$O and OH, as they are intrinsically linked through thermal dissociation and their relative shift is more significant.
We analyse the relative velocity shifts of H$_2$O and OH rather than their absolute shifts, as the absolute radial velocity of the system is affected by effects such as gravitational redshift, convective blueshift, and systematic instrumental offsets. Such uncertainties can affect the absolute inferred value of V$_\mathrm{sys}$ but would not affect the relative shift between H$_2$O and OH.

When considering both pre- and post-eclipse phases, GCMs cannot reproduce the shift between H$_2$O and OH (Fig. \ref{fig:GCM_strongdrag}). However, the pre-eclipse phases map does predict a redshift of the H$_2$O signature. 
This effect rises from the fact that the distribution of species is not constant throughout the dayside.
At the substellar point, H$_2$O is depleted and OH is enriched \citep[e.g.][]{wardenier_modelling_2023}. Meanwhile, the regions closer to the terminators are dominated by H$_2$O. In the pre-eclipse phases, the eastern terminator is visible and it has a positive relative speed (moving away from us), explaining the redshift of the H$_2$O feature.
This same reasoning can be applied to the post-eclipse phases, where the western terminator is in view. This will, however, create a blueshifted H$_2$O signal.
We find that, out of the four GCMs considered, the strong drag model has the greatest $V_{\mathrm{sys}}$ shift between molecules in the pre-eclipse phases. 
The winds that transports heat to the eastern dayside increases the OH abundance in the eastern dayside, which in turn reduces the relative velocity shift between OH and H$_2$O. Therefore, a scenario with restricted wind circulation (strong drag) should have a stronger abundance contrast in the eastern dayside creating a stronger velocity shift when observing at the pre-eclipse phases.
We note that the relative shift seen in the pre-eclipse phases is in the order of 2\,km\,s$^{-1}$, while we observe a shift of about 5\,km\,s$^{-1}$. The complex interactions in GCMs are difficult to model accurately. These hard-to-model effects include wind speed. Therefore, the absolute shifts seen in GCMs should be taken with caution and they should be looked at from a qualitative point of view.

The fact that heterogeneous distributions of species can create velocity shifts can be explained using Eq. \ref{eq:rv_on_pla}.
A difference in longitude of the signal will not directly create a shift in $V_{\mathrm{sys}}$, but instead a shift in the phase of the radial velocity curve. When varying $\theta$ in Eq. \ref{eq:rv_on_pla}, the resulting sine curve will not have the same phase as the phase of the planet around its host star. This makes sense since when the planet is at phase 0, for example, the point of (non-zero) longitude $\theta$ will either be trailing or leading the centre of the planet.
However, our dataset only covers a part of the full orbit, where the dayside is in view. 
It happens that the pre- and post-eclipse phases is when the phase shift creates a similar wavelength shift, which can be approximated by a $V_{\mathrm{sys}}$ shift.
Therefore, if we assume that the radial velocity curve of the chemical species has the same phase as the orbital movement of the planet and all observations are taken on the dayside, we should expect to see a shift in $V_{\mathrm{sys}}$, even if, in reality, this shift arises from a difference in the phase of the radial velocity curve.

If the H$_2$O signal was only coming from the eastern dayside, we would expect a $V_{\mathrm{sys}}$ shift, similar to the observations. This would also imply that the OH signal contribution is symmetrically distributed around the substellar point. However, if the H$_2$O contribution is too much to the east, we should not be able to detect the H$_2$O in the post-eclipse phases. 
We propose that the H$_2$O signal on the dayside of the planet is a gradient, with the strongest signal on the eastern part of the dayside which decreases as it approaches the western dayside. As such, the observed spectra from all angles would come predominantly from the eastern part of the visible hemisphere. This is consistent with the observation that the pre-eclipse cross-correlation map seem to have a stronger signal than the post-eclipse map (Fig. \ref{fig:kp_vsys_pre_post}).
Overall, our observations suggest that WASP-121\,b is in a strong drag regime where the H$_2$O signature is weaker on the western dayside.
This weakening of the H$_2$O signal might be caused by several effects. Firstly, it could simply be that the H$_2$O is less abundant in the western dayside.
Secondly, the vertical temperature profile might be close to isothermal in the western dayside.
Thirdly, clouds could hide the H$_2$O features. However, clouds are not expected to be on the daysides of UHJs due to the extreme temperatures \citep[e.g.][]{helling_cloud_2021, gao_aerosol_2020}, therefore we favour the two scenarios mentioned above.
Because of the correlation between temperature, abundance, and cloud-top pressure, it is impossible to distinguish between these scenarios with high resolution alone.

The strong drag scenario was used to explain observations of other UHJs.
WASP-18\,b showed a symmetrical temperature profile around the substellar point \citep{coulombe_broadband_2023}, which the authors attribute this strong drag to magnetic effects \citep{rauscher_three-dimensional_2013, beltz_exploring_2021, beltz_magnetic_2022}. UHJs atmospheres contain ions due to the thermal stripping of electrons around metallic elements. The interaction of these ions with the magnetic field of the planet causes drag that is reflected by low wind speeds and poor heat distribution in the atmosphere.

Day-to-night winds could also induce $V_{\mathrm{sys}}$ shifts. These strong winds generate from the substellar point and travel towards the nigthside, such that species that are carried by that wind are expected to be redshifted when observing the dayside. The H$_2$O might be one of these molecular species. We expect H$_2$O to be close to the terminator regions, where the day-to-night wind is strong. It is predicted that this wind could be as fast as 10\,km\,s$^{-1}$ at the mbar level, which is consistent with our observations \citep{kempton_constraining_2012}.
The OH, on the other hand, is expected to be present closer to the substellar point, due to the strong thermal dissociation of H$_2$O in this region. The air in the substellar point should rise in altitude, therefore a species in that wind should not be redshifted, but blueshifted instead \citep{costa_silva_espresso_2024}.
This might explain the small blue shift seen in the OH signal, however, this blue shift is not significant enough to draw conclusive results.
GCMs naturally include day-to-night winds. However, if our observed redshift of H2O is caused by day-to-night winds, the model is under-predicting its strength. One way to increase the day-to-night wind speeds is to decrease the drag (i.e. increase the drag timescale) in the GCM, but such a model would be at odds with previous observations of WASP-121\,b \citep[e.g.][]{mikal-evans_jwst_2023, wardenier_phase-resolving_2024}

Other works observed shifts between species in the $K_p-V_{\mathrm{sys}}$ maps \citep[e.g.][]{brogi_roasting_2023, prinoth_titanium_2022, cont_detection_2021}.
Notably, \citet{brogi_roasting_2023} found velocity shifts between H$_2$O, OH and CO in a single dayside observation of WASP-18\,b. They attributed these shifts mainly to atmospheric dynamics. They explained the difference in shift between H$_2$O and OH by the different distribution of the molecular species caused by the increased dissociation of H$_2$O into OH in the hot spot. 
This explanation is somewhat suitable for observations of only post-eclipse phases or only pre-eclipse phases, but not both. As mentioned above, we require a mechanism to weaken the H$_2$O signal in the western dayside to explain our relative shift.
Interestingly, even with the velocity shifts between molecular detections, \citet{brogi_roasting_2023} retrieved constrained abundances for all of their detected species.
This is possibly because their dataset only had one observation night. With our six nights, covering both the pre-eclipse and post-eclipse phases, the cross-correlation signal is more local and does not have an extended trail like we would expect from having only one night. Therefore, our retrieval with six nights cannot find a common point in velocity space where it can get the contribution of all chemical species with a relative shift present.

In WASP-121\,b, \citet{pelletier_crires_2024} found a blueshifted H$_2$O signal using CRIRES+ pre- and post-eclispe dayside observations. Although less significant than our observed shift, their H$_2$O signal is offset by $\sim -2.5\,$km\,s$^{-1}$ relative to other detected species, in disagreement with our redshifted feature. 
They point out that a possible explanation is that the western dayside is cooler than the eastern dayside. In this case, the H$_2$O should be underabundant in the eastern dayside due to thermal dissociation. The western dayside would contain a greater concentration of H$_2$O, which would contribute more to the signal.
Meanwhile, \citet{smith_roasting_2024} found a redshifted H$_2$O feature, along with a marginally blueshifted OH signal using pre- and post-eclipse IGRINS observations. Their relative redshift between OH and H$_2$O is $\sim 2$\,km\,s$^{-1}$, significantly lower than our observed shift. Their hypotheses to explain this shift include day-to-night winds and signal coming from a region away from the substellar point, both similar to our hypotheses.

The contradicting results between \citet{pelletier_crires_2024}, \citet{smith_roasting_2024}, and this work support the hypothesis that noise might play an important role in our cross-correlation. The relatively low S/N in our observations could introduce some additional noise source. The H$_2$O spectral lines could preferentially be redshifted due to physical, systematic, or instrumental effects.

\subsection{Metallic ratio}

Using a free retrieval, we constrain the abundance of Fe and Mg (Fig. \ref{fig:retrievalres}) to log$_{10}$Fe$= -2.91^{+1.10}_{-1.36}$ and log$_{10}$Mg$= -2.75^{+1.09}_{-1.37}$.
Albeit with large uncertainties, these abundances are marginally elevated compared to a solar atmospheric composition.
Observations and numerical simulations predict that WASP-121\,b's metallicity is super-solar with a metal enrichment of 10–30 times the solar value \citep{daylan_tess_2021, mikal-evans_emission_2019, evans_optical_2018, thorngren_connecting_2019}. 
We find that an atmospheric model with a metallicity of $\sim$30 has Fe and Mg mixing ratios that are close to the retrieved abundances. Here, the model used has a vertical temperature profile corresponding to the median retrieved profile. However, we note the high uncertainties on the retrieved abundances indicate that a large range of metallicities is consistent with our results.

\begin{figure}
    \centering
    \includegraphics[width=\linewidth]{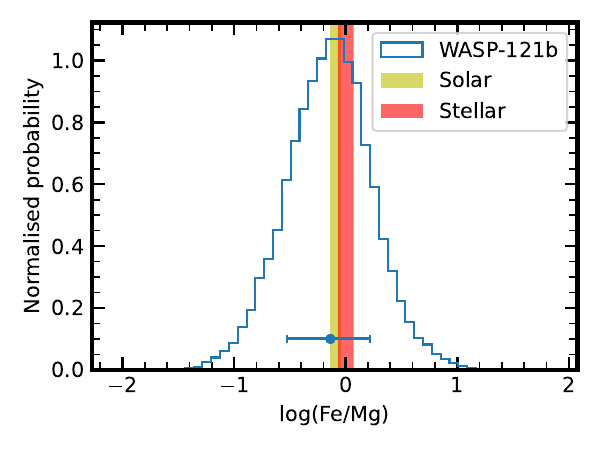}
    \caption{Posterior of the ratio of Fe over Mg in the atmosphere of WASP-121\,b from retrieval results.  The blue point is the median of the blue distribution, with the error bars being the 1\,$\sigma$ uncertainties of that same posterior.
    The yellow and red area represents the $1\,\sigma$ uncertainties of the Fe/Mg for the Sun \citep{asplund_chemical_2021} and WASP-121 \citep{polanski_chemical_2022}, respectively.
    The retrieved ratio is consistent with both the solar and stellar value.}
    \label{fig:Fe_over_Mg}
\end{figure}

The calculated abundance ratio between Fe and Mg using the retrieval results gives a value of $\log_{10}($Fe/Mg$) = -0.14^{+0.36}_{-0.39}$ (Fig. \ref{fig:Fe_over_Mg}). This is well consistent with both the solar value of $-0.09\pm0.05$ \citep{asplund_chemical_2021} and the stellar value of $0.00\pm0.07$ \citep{polanski_chemical_2022}. Our retrieval value is also consistent with the Fe/Mg retrieved by \citet{maguire_high-resolution_2023} from three different ESPRESSO transits of WASP-121\,b and by \citet{smith_roasting_2024} from two IGRINS dayside observations, which both find a log ratio close to 0. 
Our result is in agreement with planet formation theories that hot Jupiters, and planets in general, are expected to have metallic ratios that are similar to their host star \citep[e.g.][]{thiabaud_elemental_2015, adibekyan_compositional_2021, chachan_breaking_2023}. This trend has observed in other planets as well \citep[e.g.][]{dorn_bayesian_2017, pelletier_vanadium_2023}.

Although the retrieval does not constrain the absolute abundances of Fe and Mg well due to degeneracies with both the temperature structure and the continuum set here by the H$^-$ bound-free opacity \citep[][Vaulato et al. submitted]{gray_observation_2021, arcangeli_h-_2018, parmentier_exoplanet_2018}, their abundance ratio can still be precisely determined.
It has been shown that retrievals give mostly better constraints on elemental ratios rather than absolute abundance \citep[e.g.][]{brogi_retrieving_2019, line_solar_2021, gandhi_retrieval_2023, maguire_high-resolution_2023}.
Indeed, the 2D distributions (Corner plot of Fig. \ref{fig:retrievalres}) reveal positive correlations between species. Since the log of a ratio is the subtraction of the log abundances, we can imagine collapsing the 2D distribution from bottom left to top right. When there is a strong positive correlation between parameters, this will remove the axis with the highest variance, and will give a better constraint than the individual 1D distribution.

We note that the Fe and Mg signal comes primarily from one night (2024 February 2).
Several factors can explain differences in the detection strengths between observing nights. Such factors include atmospheric variability \citep{ouyang_detection_2023}, but also non-planetary effects. Given the relatively low S/N of our dataset, it is virtually impossible to distinguish which effect is primarily responsible for the observed differences.
To determine whether this single night substantially biases the posteriors, we run a retrieval without the night of 2024 February 2. The resulting retrieval is still able to constrain the Fe and Mg. However, the constraints on the abundances are wider, with a metallic ratio of $\log_{10}($Fe/Mg$) = -1.02^{+0.59}_{-1.56}$ being retrieved.
Notably this retrieval prefers less Fe relative to Mg than the retrieval where all six nights are considered, resulting in a lower Fe/Mg (but still consistent within 2\,$\sigma$).

\section{Conclusions} \label{sec:conclusion}

In this work, we present a dayside analysis of the UHJ WASP-121\,b using near-infrared high-resolution spectra taken with NIRPS. 
We report detections of H$_2$O, OH, Fe and Mg.
The simultaneous detections of H$_2$O and OH is evidence of thermal dissociation in the dayside photosphere of WASP-121\,b. This makes WASP-121\,b one of the few UHJs where OH was detected.
The atmospheric retrieval analysis returns a ratio of log$_{10}$(OH/H$_2$O) $= -0.15\pm{0.20}$ indicating that the abundance of OH is similar to that of H$_2$O in the photosphere.
This ratio is in agreement with expectation from chemical equilibrium models. However, since water dissociation is highly correlated with the pressure, the ratio varies considerably throughout the atmosphere and an uncertainty on the pressure level probed can lead to drastically different conclusions.
We also find a preliminary trend that hotter UHJs exhibit more dissociation in their atmosphere than cooler UHJs. This trend follows closely the expectations from chemical equilibrium models. These results indicate that UHJ atmospheres can be well approximated by models using only thermochemical processes.

All detection have a measured $K_p$ that is fully consistent with the expected planetary semi-amplitude considering the planet's rotational velocity. We explore two ways of calculating the expected $K_p$ and notice that their uncertainties are significantly different. We urge caution when using the value of $K_p$ without taking its uncertainty into account.

The cross-correlation maps show a significant redshift of planetary H$_2$O signature compared to other detected species. Using a retrieval approach, we quantify the shift to $4.79^{+0.93}_{-0.97}\,$km$\,$s$^{-1}$. We compare our observations to GCM predictions and find that the models cannot fully explain the observed relative shift between molecular species.
A possible explanation is that the H$_2$O signal on the western dayside is weakened by some atmospheric process, such as an under abundance of H$_2$O or an isothermal temperature profile.
Day-to-night winds where different species are trapped in winds of different regimes could be responsible for a strong shift in systemic velocity. Alternatively, systematic or instrumental noise might be responsible for the shift.

We include the abundance of Fe and Mg as free parameters in the retrieval. The posteriors on the abundance of the individual species are broad. However, the ratio of these species is well constrained with a value of $\log_{10}($Fe/Mg$) = -0.14^{+0.36}_{-0.39}$. This is consistent with the solar value and the value of the host star WASP-121 within 1\,$\sigma$.

With its wide wavelength range and high resolving power, NIRPS is able to detect OH alongside H$_2$O in the atmosphere of WASP-121\,b. Furthermore, it reveals an intriguing velocity shift between molecular species. These results are one of the first demonstrations of the power of NIRPS for atmospheric characterisation.

\begin{acknowledgements}
We express our gratitude to the referee, whose comments significantly helped improve the quality of the paper.
LB, RA, BB, JPW, NJC, \'EA, FBa, RD, LMa, AB, CC, L-PC, AD-B, LD, PL, OL, JS-A, PV \& TV  acknowledge the financial support of the FRQ-NT through the Centre de recherche en astrophysique du Qu\'ebec as well as the support from the Trottier Family Foundation and the Trottier Institute for Research on Exoplanets.\\
LB  acknowledges the support of the Natural Sciences and Engineering Research Council of Canada (NSERC).\\
RA  acknowledges the Swiss National Science Foundation (SNSF) support under the Post-Doc Mobility grant P500PT\_222212 and the support of the Institut Trottier de Recherche sur les Exoplan\`etes (IREx).\\
This work has been carried out within the framework of the NCCR PlanetS supported by the Swiss National Science Foundation under grants 51NF40\_182901 and 51NF40\_205606.\\
TF, XB, XDe, ACar \& VY  acknowledge funding from the French ANR under contract number ANR\-24\-CE49\-3397 (ORVET), and the French National Research Agency in the framework of the Investissements d'Avenir program (ANR-15-IDEX-02), through the funding of the ``Origin of Life" project of the Grenoble-Alpes University.\\
KAM  acknowledges support from the Swiss National Science Foundation (SNSF) under the Postdoc Mobility grant P500PT\_230225.\\
\'EA, FBa, RD, LMa, TA, J-SM, MO, JS-A \& PV  acknowledges support from Canada Foundation for Innovation (CFI) program, the Universit\'e de Montr\'eal and Universit\'e Laval, the Canada Economic Development (CED) program and the Ministere of Economy, Innovation and Energy (MEIE).\\
SCB, NCS, ARCS \& EC  acknowledge the support from FCT - Funda\c{c}\~ao para a Ci\^encia e a Tecnologia through national funds by these grants: UIDB/04434/2020, UIDP/04434/2020.\\
SCB   acknowledges the support from Funda\c{c}\~ao para a Ci\^encia e Tecnologia (FCT) in the form of a work contract through the Scientific Employment Incentive program with reference 2023.06687.CEECIND and DOI https://doi.org/10.54499/2023.06687.CEECIND/CP2839/CT0002.\\
The Board of Observational and Instrumental Astronomy (NAOS) at the Federal University of Rio Grande do Norte's research activities are supported by continuous grants from the Brazilian funding agency CNPq. This study was partially funded by the Coordena\c{c}\~ao de Aperfei\c{c}oamento de Pessoal de N\'ivel Superior—Brasil (CAPES) — Finance Code 001 and the CAPES-Print program.\\
BLCM \& AMM  acknowledge CAPES postdoctoral fellowships.\\
BLCM  acknowledges CNPq research fellowships (Grant No. 305804/2022-7).\\
NBC  acknowledges support from an NSERC Discovery Grant, a Canada Research Chair, and an Arthur B. McDonald Fellowship, and thanks the Trottier Space Institute for its financial support and dynamic intellectual environment.\\
DBF  acknowledges financial support from the Brazilian agency CNPq-PQ (Grant No. 305566/2021-0). Continuous grants from the Brazilian agency CNPq support the STELLAR TEAM of the Federal University of Ceara's research activities.\\
JRM  acknowledges CNPq research fellowships (Grant No. 308928/2019-9).\\
XDu  acknowledges the support from the European Research Council (ERC) under the European Union’s Horizon 2020 research and innovation programme (grant agreement SCORE No 851555) and from the Swiss National Science Foundation under the grant SPECTRE (No 200021\_215200).\\
DE  acknowledge support from the Swiss National Science Foundation for project 200021\_200726. The authors acknowledge the financial support of the SNSF.\\
JIGH, RR, ASM, FGT, JLR, NN \& AKS  acknowledge financial support from the Spanish Ministry of Science, Innovation and Universities (MICIU) projects PID2020-117493GB-I00 and PID2023-149982NB-I00.\\
ICL  acknowledges CNPq research fellowships (Grant No. 313103/2022-4).\\
CMo  acknowledges the funding from the Swiss National Science Foundation under grant 200021\_204847 “PlanetsInTime”.\\
Co-funded by the European Union (ERC, FIERCE, 101052347). Views and opinions expressed are however those of the author(s) only and do not necessarily reflect those of the European Union or the European Research Council. Neither the European Union nor the granting authority can be held responsible for them.\\
GAW is supported by a Discovery Grant from the Natural Sciences and Engineering Research Council (NSERC) of Canada.\\
0\\
This project has received funding from the European Research Council (ERC) under the European Union's Horizon 2020 research and innovation programme (project {\sc Spice Dune}, grant agreement No 947634). This material reflects only the authors' views and the Commission is not liable for any use that may be made of the information contained therein.\\
ARCS  acknowledges the support from Funda\c{c}ao para a Ci\^encia e a Tecnologia (FCT) through the fellowship 2021.07856.BD.\\
LD  acknowledges the support of the Natural Sciences and Engineering Research Council of Canada (NSERC) and from the Fonds de recherche du Qu\'ebec (FRQ) - Secteur Nature et technologies.\\
FG  acknowledges support from the Fonds de recherche du Qu\'ebec (FRQ) - Secteur Nature et technologies under file \#350366.\\
H.J.H. acknowledges funding from eSSENCE (grant number eSSENCE@LU 9:3), the Swedish National Research Council (project number 2023-05307), The Crafoord foundation and the Royal Physiographic Society of Lund, through The Fund of the Walter Gyllenberg Foundation.\\
NN  acknowledges financial support by Light Bridges S.L, Las Palmas de Gran Canaria.\\
NN acknowledges funding from Light Bridges for the Doctoral Thesis "Habitable Earth-like planets with ESPRESSO and NIRPS", in cooperation with the Instituto de Astrof\'isica de Canarias, and the use of Indefeasible Computer Rights (ICR) being commissioned at the ASTRO POC project in the Island of Tenerife, Canary Islands (Spain). The ICR-ASTRONOMY used for his research was provided by Light Bridges in cooperation with Hewlett Packard Enterprise (HPE).\\
CPi  acknowledges support from the NSERC Vanier scholarship, and the Trottier Family Foundation. CPi  also acknowledges support from the E. Margaret Burbidge Prize Postdoctoral Fellowship from the Brinson Foundation.\\
AKS  acknowledges financial support from La Caixa Foundation (ID 100010434) under the grant LCF/BQ/DI23/11990071.\\
TV  acknowledges support from the Fonds de recherche du Qu\'ebec (FRQ) - Secteur Nature et technologies under file \#320056.

\end{acknowledgements}

\bibliographystyle{aa}
\bibliography{NIRPS_WASP121b}

\begin{thebibliography}{131}
\expandafter\ifx\csname natexlab\endcsname\relax\def\natexlab#1{#1}\fi

\bibitem[{Adibekyan {et~al.}(2021)Adibekyan, Dorn, Sousa, Santos, Bitsch, Israelian, Mordasini, Barros, Delgado~Mena, Demangeon, Faria, Figueira, Hakobyan, Oshagh, Soares, Kunitomo, Takeda, Jofré, Petrucci, \& Martioli}]{adibekyan_compositional_2021}
Adibekyan, V., Dorn, C., Sousa, S.~G., {et~al.} 2021, Science, 374, 330, aDS Bibcode: 2021Sci...374..330A

\bibitem[{Allart {et~al.}(2022)Allart, Lovis, Faria, Dumusque, Sosnowska, Figueira, Silva, Mehner, Pepe, Cristiani, Rebolo, Santos, Adibekyan, Cupani, Marcantonio, D’Odorico, Hernández, Martins, Milaković, Nunes, Sozzetti, Mascareño, Tabernero, \& Osorio}]{allart_automatic_2022}
Allart, R., Lovis, C., Faria, J., {et~al.} 2022, Astronomy \& Astrophysics, 666, A196

\bibitem[{Arcangeli {et~al.}(2018)Arcangeli, Désert, Line, Bean, Parmentier, Stevenson, Kreidberg, Fortney, Mansfield, \& Showman}]{arcangeli_h-_2018}
Arcangeli, J., Désert, J.-M., Line, M.~R., {et~al.} 2018, The Astrophysical Journal, 855, L30

\bibitem[{Artigau {et~al.}(2024)Artigau, Bouchy, Doyon, Baron, Malo, Wildi, Pepe, Cook, Thibault, Reshetov, Dumusque, Lovis, Sosnowska, Martins, De~Medeiros, Delfosse, Santos, Rebolo, Abreu, Allain, Allart, Auger, Barros, Bazinet, Blind, Boisse, Bonfils, Bourrier, Bovay, Broeg, Brousseau, Bruniquel, Cabral, Cadieux, Carmona, Carteret, Challita, Chazelas, Cloutier, Coelho, Cointepas, Conod, Cowan, Cristo, da~Silva, Dauplaise, Gomes, Delgado-Mena, Ehrenreich, Faria, Figueira, Forveille, Frensch, Gagné, Genest, Genolet, Hernández, Témich, Grieves, Hernandez, Hobson, Hoeijmakers, Kerley, Krishnamurthy, Lafrenière, Lamontagne, Larue, Leaf, Leão, Lim, Curto, Martins, Melo, Messias, Mignon, Moranta, Mordasini, Moulla, Mounzer, L'Heureux, Nari, Nielsen, Osborn, Parc, Pasquini, Passegger, Pelletier, Peroux, Piaulet, Plotnykov, Poulin-Girard, Rasilla, Saint-Antoine, Sarajlic, Segovia, Seidel, Ségransan, Silva, Srivastava, Stefanov, Mascareño, Sordet, Teixeira, Udry, Valencia, Vallée, Vandal, Vaulato, Wade,
  Wardenier, Wehbé, Weisserman, Wevers, \& Zins}]{artigau_nirps_2024}
Artigau, {\'E}., Bouchy, F., Doyon, R., {et~al.} 2024, {NIRPS} first light and early science: breaking the 1 m/s {RV} precision barrier at infrared wavelengths

\bibitem[{Artigau {et~al.}(2022)Artigau, Cadieux, Cook, Doyon, Vandal, Donati, Moutou, Delfosse, Fouqué, Martioli, Bouchy, Parsons, Carmona, Dumusque, Astudillo-Defru, Bonfils, \& Mignon}]{artigau_line-by-line_2022}
Artigau, {\'E}., Cadieux, C., Cook, N.~J., {et~al.} 2022, The Astronomical Journal, 164, 84, publisher: The American Astronomical Society

\bibitem[{Asplund {et~al.}(2021)Asplund, Amarsi, \& Grevesse}]{asplund_chemical_2021}
Asplund, M., Amarsi, A.~M., \& Grevesse, N. 2021, Astronomy \& Astrophysics, 653, A141

\bibitem[{Azevedo~Silva {et~al.}(2022)Azevedo~Silva, Demangeon, Santos, Allart, Borsa, Cristo, Esparza-Borges, Seidel, Palle, Sousa, Tabernero, Osorio, Cristiani, Pepe, Rebolo, Adibekyan, Alibert, Barros, Bouchy, Bourrier, Curto, Marcantonio, D’Odorico, Ehrenreich, Figueira, Hernández, Lovis, Martins, Mehner, Micela, Molaro, Mounzer, Nunes, Sozzetti, Mascareño, \& Udry}]{azevedo_silva_detection_2022}
Azevedo~Silva, T., Demangeon, O. D.~S., Santos, N.~C., {et~al.} 2022, Astronomy \& Astrophysics, 666, L10

\bibitem[{Baeyens {et~al.}(2024)Baeyens, Désert, Petrignani, Carone, \& Schneider}]{baeyens_photodissociation_2024}
Baeyens, R., Désert, J.-M., Petrignani, A., Carone, L., \& Schneider, A.~D. 2024, Astronomy \& Astrophysics, 686, A24

\bibitem[{Barber {et~al.}(2014)Barber, Strange, Hill, Polyansky, Mellau, Yurchenko, \& Tennyson}]{barber_exomol_2014}
Barber, R.~J., Strange, J.~K., Hill, C., {et~al.} 2014, Monthly Notices of the Royal Astronomical Society, 437, 1828

\bibitem[{Bazinet {et~al.}(2024)Bazinet, Pelletier, Benneke, Salinas, \& Mace}]{bazinet_subsolar_2024}
Bazinet, L., Pelletier, S., Benneke, B., Salinas, R., \& Mace, G.~N. 2024, The Astronomical Journal, 167, 206

\bibitem[{Bell \& Cowan(2018)}]{bell_increased_2018}
Bell, T.~J. \& Cowan, N.~B. 2018, The Astrophysical Journal, 857, L20

\bibitem[{Bello-Arufe {et~al.}(2022)Bello-Arufe, Cabot, Mendonça, Buchhave, \& Rathcke}]{bello-arufe_mining_2022}
Bello-Arufe, A., Cabot, S. H.~C., Mendonça, J.~M., Buchhave, L.~A., \& Rathcke, A.~D. 2022, The Astronomical Journal, 163, 96

\bibitem[{Beltz {et~al.}(2022)Beltz, Rauscher, Kempton, Malsky, Ochs, Arora, \& Savel}]{beltz_magnetic_2022}
Beltz, H., Rauscher, E., Kempton, E. M.-R., {et~al.} 2022, The Astronomical Journal, 164, 140

\bibitem[{Beltz {et~al.}(2021)Beltz, Rauscher, Roman, \& Guilliat}]{beltz_exploring_2021}
Beltz, H., Rauscher, E., Roman, M.~T., \& Guilliat, A. 2021, The Astronomical Journal, 163, 35

\bibitem[{Ben-Yami {et~al.}(2020)Ben-Yami, Madhusudhan, Cabot, Constantinou, Piette, Gandhi, \& Welbanks}]{ben-yami_neutral_2020}
Ben-Yami, M., Madhusudhan, N., Cabot, S. H.~C., {et~al.} 2020, The Astrophysical Journal Letters, 897, L5

\bibitem[{Benneke(2015)}]{benneke_strict_2015}
Benneke, B. 2015, Strict {Upper} {Limits} on the {Carbon}-to-{Oxygen} {Ratios} of {Eight} {Hot} {Jupiters} from {Self}-{Consistent} {Atmospheric} {Retrieval}

\bibitem[{Benneke {et~al.}(2019{\natexlab{a}})Benneke, Knutson, Lothringer, Crossfield, Moses, Morley, Kreidberg, Fulton, Dragomir, Howard, Wong, Désert, McCullough, Kempton, Fortney, Gilliland, Deming, \& Kammer}]{benneke_sub-neptune_2019}
Benneke, B., Knutson, H.~A., Lothringer, J., {et~al.} 2019{\natexlab{a}}, Nature Astronomy, 3, 813

\bibitem[{Benneke \& Seager(2012)}]{benneke_atmospheric_2012}
Benneke, B. \& Seager, S. 2012, The Astrophysical Journal, 753, 100

\bibitem[{Benneke \& Seager(2013)}]{benneke_how_2013}
Benneke, B. \& Seager, S. 2013, The Astrophysical Journal, 778, 153

\bibitem[{Benneke {et~al.}(2019{\natexlab{b}})Benneke, Wong, Piaulet, Knutson, Lothringer, Morley, Crossfield, Gao, Greene, Dressing, Dragomir, Howard, McCullough, Kempton, Fortney, \& Fraine}]{benneke_water_2019}
Benneke, B., Wong, I., Piaulet, C., {et~al.} 2019{\natexlab{b}}, The Astrophysical Journal Letters, 887, L14

\bibitem[{Birkby(2018)}]{birkby_spectroscopic_2018}
Birkby, J.~L. 2018, in Handbook of {Exoplanets}, ed. H.~J. Deeg \& J.~A. Belmonte (Cham: Springer International Publishing), 1485--1508

\bibitem[{Borsa {et~al.}(2021)Borsa, Allart, Casasayas-Barris, Tabernero, Osorio, Cristiani, Pepe, Rebolo, Santos, Adibekyan, Bourrier, Demangeon, Ehrenreich, Pallé, Sousa, Lillo-Box, Lovis, Micela, Oshagh, Poretti, Sozzetti, Prieto, Alibert, Amate, Benz, Bouchy, Cabral, Dekker, D’Odorico, Marcantonio, Figueira, Santos, Hernández, Curto, Manescau, Martins, Mégevand, Mehner, Molaro, Nunes, Riva, Mascareño, Udry, \& Zerbi}]{borsa_atmospheric_2021}
Borsa, F., Allart, R., Casasayas-Barris, N., {et~al.} 2021, Astronomy \& Astrophysics, 645, A24

\bibitem[{Borysow(2002)}]{borysow_collision-induced_2002}
Borysow, A. 2002, Astronomy \& Astrophysics, 390, 779

\bibitem[{Bouchy {et~al.}(2017)Bouchy, Doyon, Artigau, Melo, Hernandez, Wildi, Delfosse, Lovis, Figueira, Canto~Martins, González~Hernández, Thibault, Reshetov, Pepe, Santos, de~Medeiros, Rebolo, Abreu, Adibekyan, Bandy, Benz, Blind, Bohlender, Boisse, Bovay, Broeg, Brousseau, Cabral, Chazelas, Cloutier, Coelho, Conod, Cumming, Delabre, Genolet, Hagelberg, Jayawardhana, Käufl, Lafrenière, de~Castro~Leão, Malo, de~Medeiros~Martins, Matthews, Metchev, Oshagh, Ouellet, Parro, Rasilla~Piñeiro, Santos, Sarajlic, Segovia, Sordet, Udry, Valencia, Vallée, Venn, Wade, \& Saddlemyer}]{bouchy_near-infrared_2017}
Bouchy, F., Doyon, R., Artigau, {\'E}., {et~al.} 2017, The Messenger, 169, 21, aDS Bibcode: 2017Msngr.169...21B

\bibitem[{Bourrier {et~al.}(2020)Bourrier, Ehrenreich, Lendl, Cretignier, Allart, Dumusque, Cegla, Suárez-Mascareño, Wyttenbach, Hoeijmakers, Melo, Kuntzer, Astudillo-Defru, Giles, Heng, Kitzmann, Lavie, Lovis, Murgas, Nascimbeni, Pepe, Pino, Segransan, \& Udry}]{bourrier_hot_2020}
Bourrier, V., Ehrenreich, D., Lendl, M., {et~al.} 2020, Astronomy \& Astrophysics, 635, A205

\bibitem[{Brogi {et~al.}(2023)Brogi, Emeka-Okafor, Line, Gandhi, Pino, Kempton, Rauscher, Parmentier, Bean, Mace, Cowan, Shkolnik, Wardenier, Mansfield, Welbanks, Smith, Fortney, Birkby, Zalesky, Dang, Patience, \& Désert}]{brogi_roasting_2023}
Brogi, M., Emeka-Okafor, V., Line, M.~R., {et~al.} 2023, The Astronomical Journal, 165, 91

\bibitem[{Brogi {et~al.}(2016)Brogi, Kok, Albrecht, Snellen, Birkby, \& Schwarz}]{brogi_rotation_2016}
Brogi, M., Kok, R. J.~d., Albrecht, S., {et~al.} 2016, The Astrophysical Journal, 817, 106

\bibitem[{Brogi \& Line(2019)}]{brogi_retrieving_2019}
Brogi, M. \& Line, M.~R. 2019, The Astronomical Journal, 157, 114

\bibitem[{Brogi {et~al.}(2012)Brogi, Snellen, de~Kok, Albrecht, Birkby, \& de~Mooij}]{brogi_signature_2012}
Brogi, M., Snellen, I. A.~G., de~Kok, R.~J., {et~al.} 2012, Nature, 486, 502

\bibitem[{Brown {et~al.}(2018)Brown, Vallenari, Prusti, Bruijne, Babusiaux, Bailer-Jones, Biermann, Evans, Eyer, Jansen, Jordi, Klioner, Lammers, Lindegren, Luri, Mignard, Panem, Pourbaix, Randich, Sartoretti, Siddiqui, Soubiran, Leeuwen, Walton, Arenou, Bastian, Cropper, Drimmel, Katz, Lattanzi, Bakker, Cacciari, Castañeda, Chaoul, Cheek, Angeli, Fabricius, Guerra, Holl, Masana, Messineo, Mowlavi, Nienartowicz, Panuzzo, Portell, Riello, Seabroke, Tanga, Thévenin, Gracia-Abril, Comoretto, Garcia-Reinaldos, Teyssier, Altmann, Andrae, Audard, Bellas-Velidis, Benson, Berthier, Blomme, Burgess, Busso, Carry, Cellino, Clementini, Clotet, Creevey, Davidson, Ridder, Delchambre, Dell’Oro, Ducourant, Fernández-Hernández, Fouesneau, Frémat, Galluccio, García-Torres, González-Núñez, González-Vidal, Gosset, Guy, Halbwachs, Hambly, Harrison, Hernández, Hestroffer, Hodgkin, Hutton, Jasniewicz, Jean-Antoine-Piccolo, Jordan, Korn, Krone-Martins, Lanzafame, Lebzelter, Löffler, Manteiga, Marrese, Martín-Fleitas,
  Moitinho, Mora, Muinonen, Osinde, Pancino, Pauwels, Petit, Recio-Blanco, Richards, Rimoldini, Robin, Sarro, Siopis, Smith, Sozzetti, Süveges, Torra, Reeven, Abbas, Aramburu, Accart, Aerts, Altavilla, Álvarez, Alvarez, Alves, Anderson, Andrei, Varela, Antiche, Antoja, Arcay, Astraatmadja, Bach, Baker, Balaguer-Núñez, Balm, Barache, Barata, Barbato, Barblan, Barklem, Barrado, Barros, Barstow, Muñoz, Bassilana, Becciani, Bellazzini, Berihuete, Bertone, Bianchi, Bienaymé, Blanco-Cuaresma, Boch, Boeche, Bombrun, Borrachero, Bossini, Bouquillon, Bourda, Bragaglia, Bramante, Breddels, Bressan, Brouillet, Brüsemeister, Brugaletta, Bucciarelli, Burlacu, Busonero, Butkevich, Buzzi, Caffau, Cancelliere, Cannizzaro, Cantat-Gaudin, Carballo, Carlucci, Carrasco, Casamiquela, Castellani, Castro-Ginard, Charlot, Chemin, Chiavassa, Cocozza, Costigan, Cowell, Crifo, Crosta, Crowley, Cuypers†, Dafonte, Damerdji, Dapergolas, David, David, Laverny, Luise, March, Martino, Souza, Torres, Debosscher, Pozo, Delbo, Delgado,
  Delgado, Matteo, Diakite, Diener, Distefano, Dolding, Drazinos, Durán, Edvardsson, Enke, Eriksson, Esquej, Bontemps, Fabre, Fabrizio, Faigler, Falcão, Casas, Federici, Fedorets, Fernique, Figueras, Filippi, Findeisen, Fonti, Fraile, Fraser, Frézouls, Gai, Galleti, Garabato, García-Sedano, Garofalo, Garralda, Gavel, Gavras, Gerssen, Geyer, Giacobbe, Gilmore, Girona, Giuffrida, Glass, Gomes, Granvik, Gueguen, Guerrier, Guiraud, Gutiérrez-Sánchez, Haigron, Hatzidimitriou, Hauser, Haywood, Heiter, Helmi, Heu, Hilger, Hobbs, Hofmann, Holland, Huckle, Hypki, Icardi, Janßen, Fombelle, Jonker, Juhász, Julbe, Karampelas, Kewley, Klar, Kochoska, Kohley, Kolenberg, Kontizas, Kontizas, Koposov, Kordopatis, Kostrzewa-Rutkowska, Koubsky, Lambert, Lanza, Lasne, Lavigne, Fustec, Poncin-Lafitte, Lebreton, Leccia, Leclerc, Lecoeur-Taibi, Lenhardt, Leroux, Liao, Licata, Lindstrøm, Lister, Livanou, Lobel, López, Managau, Mann, Mantelet, Marchal, Marchant, Marconi, Marinoni, Marschalkó, Marshall, Martino, Marton,
  Mary, Massari, Matijevič, Mazeh, McMillan, Messina, Michalik, Millar, Molina, Molinaro, Molnár, Montegriffo, Mor, Morbidelli, Morel, Morris, Mulone, Muraveva, Musella, Nelemans, Nicastro, Noval, O’Mullane, Ordénovic, Ordóñez-Blanco, Osborne, Pagani, Pagano, Pailler, Palacin, Palaversa, Panahi, Pawlak, Piersimoni, Pineau, Plachy, Plum, Poggio, Poujoulet, Prša, Pulone, Racero, Ragaini, Rambaux, Ramos-Lerate, Regibo, Reylé, Riclet, Ripepi, Riva, Rivard, Rixon, Roegiers, Roelens, Romero-Gómez, Rowell, Royer, Ruiz-Dern, Sadowski, Sellés, Sahlmann, Salgado, Salguero, Sanna, Santana-Ros, Sarasso, Savietto, Schultheis, Sciacca, Segol, Segovia, Ségransan, Shih, Siltala, Silva, Smart, Smith, Solano, Solitro, Sordo, Nieto, Souchay, Spagna, Spoto, Stampa, Steele, Steidelmüller, Stephenson, Stoev, Suess, Surdej, Szabados, Szegedi-Elek, Tapiador, Taris, Tauran, Taylor, Teixeira, Terrett, Teyssandier, Thuillot, Titarenko, Clotet, Turon, Ulla, Utrilla, Uzzi, Vaillant, Valentini, Valette, Elteren, Hemelryck,
  Leeuwen, Vaschetto, Vecchiato, Veljanoski, Viala, Vicente, Vogt, Essen, Voss, Votruba, Voutsinas, Walmsley, Weiler, Wertz, Wevers, Wyrzykowski, Yoldas, Žerjal, Ziaeepour, Zorec, Zschocke, Zucker, Zurbach, \& Zwitter}]{brown_gaia_2018}
Brown, A. G.~A., Vallenari, A., Prusti, T., {et~al.} 2018, Astronomy \& Astrophysics, 616, A1

\bibitem[{Chachan {et~al.}(2023)Chachan, Knutson, Lothringer, \& Blake}]{chachan_breaking_2023}
Chachan, Y., Knutson, H.~A., Lothringer, J., \& Blake, G.~A. 2023, The Astrophysical Journal, 943, 112

\bibitem[{Changeat {et~al.}(2024)Changeat, Skinner, Cho, Nättilä, Waldmann, Al-Refaie, Dyrek, Edwards, Mikal-Evans, Joshua, Morello, Skaf, Tsiaras, Venot, \& Yip}]{changeat_is_2024}
Changeat, Q., Skinner, J.~W., Cho, J. Y.-K., {et~al.} 2024, The Astrophysical Journal Supplement Series, 270, 34

\bibitem[{Cheverall {et~al.}(2023)Cheverall, Madhusudhan, \& Holmberg}]{cheverall_robustness_2023}
Cheverall, C.~J., Madhusudhan, N., \& Holmberg, M. 2023, Monthly Notices of the Royal Astronomical Society, 522, 661

\bibitem[{Chubb {et~al.}(2020)Chubb, Tennyson, \& Yurchenko}]{chubb_exomol_2020}
Chubb, K.~L., Tennyson, J., \& Yurchenko, S.~N. 2020, Monthly Notices of the Royal Astronomical Society, 493, 1531

\bibitem[{Collier~Cameron {et~al.}(2010)Collier~Cameron, Guenther, Smalley, McDonald, Hebb, Andersen, Augusteijn, Barros, Brown, Cochran, Endl, Fossey, Hartmann, Maxted, Pollacco, Skillen, Telting, Waldmann, \& West}]{collier_cameron_line-profile_2010}
Collier~Cameron, A., Guenther, E., Smalley, B., {et~al.} 2010, Monthly Notices of the Royal Astronomical Society, 407, 507

\bibitem[{Cont {et~al.}(2021)Cont, Yan, Reiners, Casasayas-Barris, Mollière, Pallé, Henning, Nortmann, Stangret, Czesla, López-Puertas, Sánchez-López, Rodler, Ribas, Quirrenbach, Caballero, Amado, Carone, Khaimova, Kreidberg, Molaverdikhani, Montes, Morello, Nagel, Oshagh, \& Zechmeister}]{cont_detection_2021}
Cont, D., Yan, F., Reiners, A., {et~al.} 2021, Astronomy \& Astrophysics, 651, A33

\bibitem[{Cook {et~al.}(2022)Cook, Artigau, Doyon, Hobson, Martioli, Bouchy, Moutou, Carmona, Usher, Fouqué, Arnold, Delfosse, Boisse, Cadieux, Vandal, Donati, \& Deslières}]{cook_apero_2022}
Cook, N.~J., Artigau, {\'E}., Doyon, R., {et~al.} 2022, Publications of the Astronomical Society of the Pacific, 134, 114509

\bibitem[{Costa~Silva {et~al.}(2024)Costa~Silva, Demangeon, Santos, Ehrenreich, Lovis, Chakraborty, Lendl, Pepe, Cristiani, Rebolo, Zapatero-Osorio, Adibekyan, Alibert, Allart, Prieto, Silva, Borsa, Bourrier, Cristo, Di~Marcantonio, Esparza-Borges, Figueira, Hernández, Herrero-Cisneros, Curto, Martins, Mehner, Nunes, Palle, Pelletier, Seidel, Silva, Sousa, Sozzetti, Steiner, Mascareño, \& Udry}]{costa_silva_espresso_2024}
Costa~Silva, A.~R., Demangeon, O. D.~S., Santos, N.~C., {et~al.} 2024, Astronomy \& Astrophysics, 689, A8

\bibitem[{Coulombe {et~al.}(2023)Coulombe, Benneke, Challener, Piette, Wiser, Mansfield, MacDonald, Beltz, Feinstein, Radica, Savel, Dos~Santos, Bean, Parmentier, Wong, Rauscher, Komacek, Kempton, Tan, Hammond, Lewis, Line, Lee, Shivkumar, Crossfield, Nixon, Rackham, Wakeford, Welbanks, Zhang, Batalha, Berta-Thompson, Changeat, Désert, Espinoza, Goyal, Harrington, Knutson, Kreidberg, López-Morales, Shporer, Sing, Stevenson, Aggarwal, Ahrer, Alam, Bell, Blecic, Caceres, Carter, Casewell, Crouzet, Cubillos, Decin, Fortney, Gibson, Heng, Henning, Iro, Kendrew, Lagage, Leconte, Lendl, Lothringer, Mancini, Mikal-Evans, Molaverdikhani, Nikolov, Ohno, Palle, Piaulet, Redfield, Roy, Tsai, Venot, \& Wheatley}]{coulombe_broadband_2023}
Coulombe, L.-P., Benneke, B., Challener, R., {et~al.} 2023, Nature, 620, 292

\bibitem[{Dang {et~al.}(2024)Dang, Bell, {Ying}, {Shu}, Cowan, Bean, Deming, Kempton, Mansfield, Rauscher, Parmentier, Stevenson, Swain, Kreidberg, Kataria, Désert, Zellem, Fortney, Lewis, Line, Morley, \& Showman}]{dang_comprehensive_2024}
Dang, L., Bell, T.~J., {Ying}, {et~al.} 2024, A {Comprehensive} {Analysis} {Spitzer} 4.5 \{\vphantom{\}}{\textbackslash}textbackslashmu{\textbackslash}m {Phase} {Curve} of {Hot} {Jupiters}

\bibitem[{Daylan {et~al.}(2021)Daylan, Günther, Mikal-Evans, Sing, Wong, Shporer, Niraula, Wit, Koll, Parmentier, Fetherolf, Kane, Ricker, Vanderspek, Seager, Winn, Jenkins, Caldwell, Charbonneau, Henze, Paegert, Rinehart, Rose, Sha, Quintana, \& Villasenor}]{daylan_tess_2021}
Daylan, T., Günther, M.~N., Mikal-Evans, T., {et~al.} 2021, The Astronomical Journal, 161, 131

\bibitem[{Delrez {et~al.}(2016)Delrez, Santerne, Almenara, Anderson, Collier-Cameron, Díaz, Gillon, Hellier, Jehin, Lendl, Maxted, Neveu-VanMalle, Pepe, Pollacco, Queloz, Ségransan, Smalley, Smith, Triaud, Udry, Van~Grootel, \& West}]{delrez_wasp-121_2016}
Delrez, L., Santerne, A., Almenara, J.-M., {et~al.} 2016, Monthly Notices of the Royal Astronomical Society, 458, 4025

\bibitem[{Donati {et~al.}(2020)Donati, Kouach, Moutou, Doyon, Delfosse, Artigau, Baratchart, Lacombe, Barrick, Hébrard, Bouchy, Saddlemyer, Parès, Rabou, Micheau, Dolon, Reshetov, Challita, Carmona, Striebig, Thibault, Martioli, Cook, Fouqué, Vermeulen, Wang, Arnold, Pepe, Boisse, Figueira, Bouvier, Ray, Feugeade, Morin, Alencar, Hobson, Castilho, Udry, Santos, Hernandez, Benedict, Vallée, Gallou, Dupieux, Larrieu, Perruchot, Sottile, Moreau, Usher, Baril, Wildi, Chazelas, Malo, Bonfils, Loop, Kerley, Wevers, Dunn, Pazder, Macdonald, Dubois, Carrié, Valentin, Henault, Yan, \& Steinmetz}]{donati_spirou_2020}
Donati, J.-F., Kouach, D., Moutou, C., {et~al.} 2020, Monthly Notices of the Royal Astronomical Society, 498, 5684

\bibitem[{Dorn {et~al.}(2017)Dorn, Hinkel, \& Venturini}]{dorn_bayesian_2017}
Dorn, C., Hinkel, N.~R., \& Venturini, J. 2017, Astronomy and Astrophysics, 597, A38, publisher: EDP ADS Bibcode: 2017A\&A...597A..38D

\bibitem[{Ehrenreich {et~al.}(2020)Ehrenreich, Lovis, Allart, Zapatero~Osorio, Pepe, Cristiani, Rebolo, Santos, Borsa, Demangeon, Dumusque, González~Hernández, Casasayas-Barris, Ségransan, Sousa, Abreu, Adibekyan, Affolter, Allende~Prieto, Alibert, Aliverti, Alves, Amate, Avila, Baldini, Bandy, Benz, Bianco, Bolmont, Bouchy, Bourrier, Broeg, Cabral, Calderone, Pallé, Cegla, Cirami, Coelho, Conconi, Coretti, Cumani, Cupani, Dekker, Delabre, Deiries, D’Odorico, Di~Marcantonio, Figueira, Fragoso, Genolet, Genoni, Génova~Santos, Hara, Hughes, Iwert, Kerber, Knudstrup, Landoni, Lavie, Lizon, Lendl, Lo~Curto, Maire, Manescau, Martins, Mégevand, Mehner, Micela, Modigliani, Molaro, Monteiro, Monteiro, Moschetti, Müller, Nunes, Oggioni, Oliveira, Pariani, Pasquini, Poretti, Rasilla, Redaelli, Riva, Santana~Tschudi, Santin, Santos, Segovia~Milla, Seidel, Sosnowska, Sozzetti, Spanò, Suárez~Mascareño, Tabernero, Tenegi, Udry, Zanutta, \& Zerbi}]{ehrenreich_nightside_2020}
Ehrenreich, D., Lovis, C., Allart, R., {et~al.} 2020, Nature, 580, 597

\bibitem[{Evans {et~al.}(2018)Evans, Sing, Goyal, Nikolov, Marley, Zahnle, Henry, Barstow, Alam, Sanz-Forcada, Kataria, Lewis, Lavvas, Ballester, Ben-Jaffel, Blumenthal, Bourrier, Drummond, Muñoz, López-Morales, Tremblin, Ehrenreich, Wakeford, Buchhave, Etangs, Hébrard, \& Williamson}]{evans_optical_2018}
Evans, T.~M., Sing, D.~K., Goyal, J.~M., {et~al.} 2018, The Astronomical Journal, 156, 283

\bibitem[{Evans {et~al.}(2017)Evans, Sing, Kataria, Goyal, Nikolov, Wakeford, Deming, Marley, Amundsen, Ballester, Barstow, Ben-Jaffel, Bourrier, Buchhave, Cohen, Ehrenreich, García~Muñoz, Henry, Knutson, Lavvas, Etangs, Lewis, López-Morales, Mandell, Sanz-Forcada, Tremblin, \& Lupu}]{evans_ultrahot_2017}
Evans, T.~M., Sing, D.~K., Kataria, T., {et~al.} 2017, Nature, 548, 58

\bibitem[{Evans-Soma {et~al.}(2025)Evans-Soma, Sing, Barstow, Piette, Taylor, Lothringer, Reggiani, Goyal, Ahrer, Mayne, Rustamkulov, Kataria, Christie, Gapp, Dong, Foreman-Mackey, Hattori, \& Marley}]{evans-soma_sio_2025}
Evans-Soma, T.~M., Sing, D.~K., Barstow, J.~K., {et~al.} 2025, Nature Astronomy, 1

\bibitem[{Finnerty {et~al.}(2023)Finnerty, Schofield, Sappey, Xuan, Ruffio, Wang, Delorme, Blake, Buzard, Fitzgerald, Baker, Bartos, Bond, Calvin, Cetre, Doppmann, Echeverri, Jovanovic, Liberman, López, Martin, Mawet, Morris, Pezzato, Phillips, Ragland, Skemer, Venenciano, Wallace, Wallack, Wang, \& Wizinowich}]{finnerty_keck_2023}
Finnerty, L., Schofield, T., Sappey, B., {et~al.} 2023, The Astronomical Journal, 166, 31

\bibitem[{Foreman-Mackey {et~al.}(2013)Foreman-Mackey, Hogg, Lang, \& Goodman}]{foreman-mackey_emcee_2013}
Foreman-Mackey, D., Hogg, D.~W., Lang, D., \& Goodman, J. 2013, Publications of the Astronomical Society of the Pacific, 125, 306

\bibitem[{Fortney {et~al.}(2008)Fortney, Lodders, Marley, \& Freedman}]{fortney_unified_2008}
Fortney, J.~J., Lodders, K., Marley, M.~S., \& Freedman, R.~S. 2008, The Astrophysical Journal, 678, 1419

\bibitem[{Gandhi {et~al.}(2023)Gandhi, Kesseli, Zhang, Louca, Snellen, Brogi, Miguel, Casasayas-Barris, Pelletier, Landman, Maguire, \& Gibson}]{gandhi_retrieval_2023}
Gandhi, S., Kesseli, A., Zhang, Y., {et~al.} 2023, The Astronomical Journal, 165, 242

\bibitem[{Gandhi {et~al.}(2024)Gandhi, Landman, Snellen, Welbanks, Madhusudhan, \& Brogi}]{gandhi_revealing_2024}
Gandhi, S., Landman, R., Snellen, I., {et~al.} 2024, Monthly Notices of the Royal Astronomical Society, 530, 2885

\bibitem[{Gandhi \& Madhusudhan(2019)}]{gandhi_new_2019}
Gandhi, S. \& Madhusudhan, N. 2019, Monthly Notices of the Royal Astronomical Society, 485, 5817

\bibitem[{Gao {et~al.}(2020)Gao, Thorngren, Lee, Fortney, Morley, Wakeford, Powell, Stevenson, \& Zhang}]{gao_aerosol_2020}
Gao, P., Thorngren, D.~P., Lee, E. K.~H., {et~al.} 2020, Nature Astronomy, 4, 951, aDS Bibcode: 2020NatAs...4..951G

\bibitem[{Gibson {et~al.}(2020)Gibson, Merritt, Nugroho, Cubillos, de~Mooij, Mikal-Evans, Fossati, Lothringer, Nikolov, Sing, Spake, Watson, \& Wilson}]{gibson_detection_2020}
Gibson, N.~P., Merritt, S., Nugroho, S.~K., {et~al.} 2020, Monthly Notices of the Royal Astronomical Society, 493, 2215

\bibitem[{Gibson {et~al.}(2022)Gibson, Nugroho, Lothringer, Maguire, \& Sing}]{gibson_relative_2022}
Gibson, N.~P., Nugroho, S.~K., Lothringer, J., Maguire, C., \& Sing, D.~K. 2022, Monthly Notices of the Royal Astronomical Society, 512, 4618

\bibitem[{Gray(2021)}]{gray_observation_2021}
Gray, D.~F. 2021, The {Observation} and {Analysis} of {Stellar} {Photospheres}, iSBN: 9781009082136 Publisher: Cambridge University Press

\bibitem[{Harada {et~al.}(2021)Harada, Kempton, Rauscher, Roman, Malsky, Brinjikji, \& DiTomasso}]{harada_signatures_2021}
Harada, C.~K., Kempton, E. M.-R., Rauscher, E., {et~al.} 2021, The Astrophysical Journal, 909, 85

\bibitem[{Hargreaves {et~al.}(2020)Hargreaves, Gordon, Rey, Nikitin, Tyuterev, Kochanov, \& Rothman}]{hargreaves_accurate_2020}
Hargreaves, R.~J., Gordon, I.~E., Rey, M., {et~al.} 2020, The Astrophysical Journal Supplement Series, 247, 55

\bibitem[{Harris {et~al.}(2006)Harris, Tennyson, Kaminsky, Pavlenko, \& Jones}]{harris_improved_2006}
Harris, G.~J., Tennyson, J., Kaminsky, B.~M., Pavlenko, Y.~V., \& Jones, H. R.~A. 2006, Monthly Notices of the Royal Astronomical Society, 367, 400

\bibitem[{Hellier {et~al.}(2009)Hellier, Anderson, Cameron, Gillon, Hebb, Maxted, Queloz, Smalley, Triaud, West, Wilson, Bentley, Enoch, Horne, Irwin, Lister, Mayor, Parley, Pepe, Pollacco, Segransan, Udry, \& Wheatley}]{hellier_orbital_2009}
Hellier, C., Anderson, D.~R., Cameron, A.~C., {et~al.} 2009, Nature, 460, 1098, publisher: Nature Publishing Group

\bibitem[{Helling {et~al.}(2021)Helling, Lewis, Samra, Carone, Graham, Herbort, Chubb, Min, Waters, Parmentier, \& Mayne}]{helling_cloud_2021}
Helling, C., Lewis, D., Samra, D., {et~al.} 2021, Astronomy \& Astrophysics, 649, A44

\bibitem[{Hoeijmakers {et~al.}(2024)Hoeijmakers, Kitzmann, Morris, Prinoth, Borsato, Thorsbro, Pino, Lee, Akın, Seidel, Birkby, Allart, \& Heng}]{hoeijmakers_mantis_2024}
Hoeijmakers, H.~J., Kitzmann, D., Morris, B.~M., {et~al.} 2024, The {Mantis} {Network} {IV}: {A} titanium cold-trap on the ultra-hot {Jupiter} {WASP}-121 b

\bibitem[{Hoeijmakers {et~al.}(2020)Hoeijmakers, Seidel, Pino, Kitzmann, Sindel, Ehrenreich, Oza, Bourrier, Allart, Gebek, Lovis, Yurchenko, Astudillo-Defru, Bayliss, Cegla, Lavie, Lendl, Melo, Murgas, Nascimbeni, Pepe, Ségransan, Udry, Wyttenbach, \& Heng}]{hoeijmakers_hot_2020}
Hoeijmakers, H.~J., Seidel, J.~V., Pino, L., {et~al.} 2020, Astronomy \& Astrophysics, 641, A123

\bibitem[{Holmberg \& Madhusudhan(2022)}]{holmberg_first_2022}
Holmberg, M. \& Madhusudhan, N. 2022, The Astronomical Journal, 164, 79

\bibitem[{Kempton \& Rauscher(2012)}]{kempton_constraining_2012}
Kempton, E. M.-R. \& Rauscher, E. 2012, The Astrophysical Journal, 751, 117, publisher: The American Astronomical Society

\bibitem[{Kesseli {et~al.}(2024)Kesseli, Beltz, Rauscher, \& Snellen}]{kesseli_up_2024}
Kesseli, A.~Y., Beltz, H., Rauscher, E., \& Snellen, I. A.~G. 2024, The Astrophysical Journal, 975, 9, publisher: The American Astronomical Society

\bibitem[{Kitzmann {et~al.}(2018)Kitzmann, Heng, Rimmer, Hoeijmakers, Tsai, Malik, Lendl, Deitrick, \& Demory}]{kitzmann_peculiar_2018}
Kitzmann, D., Heng, K., Rimmer, P.~B., {et~al.} 2018, The Astrophysical Journal, 863, 183, publisher: The American Astronomical Society

\bibitem[{Kramida {et~al.}(2018)Kramida, Ralchenko, Nave, \& Reader}]{kramida_current_2018}
Kramida, A., Ralchenko, Y., Nave, G., \& Reader, J. 2018, 2018, M01.004, conference Name: APS Division of Atomic, Molecular and Optical Physics Meeting Abstracts ADS Bibcode: 2018APS..DMPM01004K

\bibitem[{Kurucz(2018)}]{kurucz_including_2018}
Kurucz, R.~L. 2018, 515, 47, conference Name: Workshop on Astrophysical Opacities ADS Bibcode: 2018ASPC..515...47K

\bibitem[{Landman {et~al.}(2021)Landman, Sánchez-López, Mollière, Kesseli, Louca, \& Snellen}]{landman_detection_2021}
Landman, R., Sánchez-López, A., Mollière, P., {et~al.} 2021, Astronomy \& Astrophysics, 656, A119

\bibitem[{Lee {et~al.}(2022{\natexlab{a}})Lee, Prinoth, Kitzmann, Tsai, Hoeijmakers, Borsato, \& Heng}]{lee_mantis_2022}
Lee, E. K.~H., Prinoth, B., Kitzmann, D., {et~al.} 2022{\natexlab{a}}, Monthly Notices of the Royal Astronomical Society, 517, 240

\bibitem[{Lee {et~al.}(2022{\natexlab{b}})Lee, Wardenier, Prinoth, Parmentier, Grimm, Baeyens, Carone, Christie, Deitrick, Kitzmann, Mayne, Roman, \& Thorsbro}]{lee_3d_2022}
Lee, E. K.~H., Wardenier, J.~P., Prinoth, B., {et~al.} 2022{\natexlab{b}}, The Astrophysical Journal, 929, 180

\bibitem[{Li {et~al.}(2015)Li, Gordon, Rothman, Tan, Hu, Kassi, Campargue, \& Medvedev}]{li_rovibrational_2015}
Li, G., Gordon, I.~E., Rothman, L.~S., {et~al.} 2015, The Astrophysical Journal Supplement Series, 216, 15

\bibitem[{Liang {et~al.}(2003)Liang, Parkinson, Lee, Yung, \& Seager}]{liang_source_2003}
Liang, M.-C., Parkinson, C.~D., Lee, A. Y.-T., Yung, Y.~L., \& Seager, S. 2003, The Astrophysical Journal, 596, L247

\bibitem[{Line {et~al.}(2021)Line, Brogi, Bean, Gandhi, Zalesky, Parmentier, Smith, Mace, Mansfield, Kempton, Fortney, Shkolnik, Patience, Rauscher, Désert, \& Wardenier}]{line_solar_2021}
Line, M.~R., Brogi, M., Bean, J.~L., {et~al.} 2021, Nature, 598, 580

\bibitem[{Louden \& Wheatley(2015)}]{louden_spatially_2015}
Louden, T. \& Wheatley, P.~J. 2015, The Astrophysical Journal Letters, 814, L24

\bibitem[{Madhusudhan \& Seager(2010)}]{madhusudhan_inference_2010}
Madhusudhan, N. \& Seager, S. 2010, The Astrophysical Journal, 725, 261

\bibitem[{Maguire {et~al.}(2023)Maguire, Gibson, Nugroho, Ramkumar, Fortune, Merritt, \& de~Mooij}]{maguire_high-resolution_2023}
Maguire, C., Gibson, N.~P., Nugroho, S.~K., {et~al.} 2023, Monthly Notices of the Royal Astronomical Society, 519, 1030

\bibitem[{Malsky {et~al.}(2021)Malsky, Rauscher, Kempton, Roman, Long, \& Harada}]{malsky_modeling_2021}
Malsky, I., Rauscher, E., Kempton, E. M.-R., {et~al.} 2021, The Astrophysical Journal, 923, 62

\bibitem[{Mansfield {et~al.}(2024)Mansfield, Line, Wardenier, Brogi, Bean, Beltz, Smith, Zalesky, Batalha, Kempton, Montet, Owen, Plavchan, \& Rauscher}]{mansfield_metallicity_2024}
Mansfield, M.~W., Line, M.~R., Wardenier, J.~P., {et~al.} 2024, The Astronomical Journal, 168, 14, publisher: The American Astronomical Society

\bibitem[{Mayor {et~al.}(2003)Mayor, Pepe, Queloz, Bouchy, Rupprecht, Lo~Curto, Avila, Benz, Bertaux, Bonfils, Dall, Dekker, Delabre, Eckert, Fleury, Gilliotte, Gojak, Guzman, Kohler, Lizon, Longinotti, Lovis, Megevand, Pasquini, Reyes, Sivan, Sosnowska, Soto, Udry, van Kesteren, Weber, \& Weilenmann}]{mayor_setting_2003}
Mayor, M., Pepe, F., Queloz, D., {et~al.} 2003, The Messenger, 114, 20, aDS Bibcode: 2003Msngr.114...20M

\bibitem[{McKemmish {et~al.}(2019)McKemmish, Masseron, Hoeijmakers, Pérez-Mesa, Grimm, Yurchenko, \& Tennyson}]{mckemmish_exomol_2019}
McKemmish, L.~K., Masseron, T., Hoeijmakers, H.~J., {et~al.} 2019, Monthly Notices of the Royal Astronomical Society, 488, 2836

\bibitem[{Merritt {et~al.}(2021)Merritt, Gibson, Nugroho, de~Mooij, Hooton, Lothringer, Matthews, Mikal-Evans, Nikolov, Sing, \& Watson}]{merritt_inventory_2021}
Merritt, S.~R., Gibson, N.~P., Nugroho, S.~K., {et~al.} 2021, Monthly Notices of the Royal Astronomical Society, 506, 3853

\bibitem[{Mikal-Evans {et~al.}(2022)Mikal-Evans, Sing, Barstow, Kataria, Goyal, Lewis, Taylor, Mayne, Daylan, Wakeford, Marley, \& Spake}]{mikal-evans_diurnal_2022}
Mikal-Evans, T., Sing, D.~K., Barstow, J.~K., {et~al.} 2022, Nature Astronomy, 6, 471

\bibitem[{Mikal-Evans {et~al.}(2023)Mikal-Evans, Sing, Dong, Foreman-Mackey, Kataria, Barstow, Goyal, Lewis, Lothringer, Mayne, Wakeford, Christie, \& Rustamkulov}]{mikal-evans_jwst_2023}
Mikal-Evans, T., Sing, D.~K., Dong, J., {et~al.} 2023, The Astrophysical Journal Letters, 943, L17

\bibitem[{Mikal-Evans {et~al.}(2019)Mikal-Evans, Sing, Goyal, Drummond, Carter, Henry, Wakeford, Lewis, Marley, Tremblin, Nikolov, Kataria, Deming, \& Ballester}]{mikal-evans_emission_2019}
Mikal-Evans, T., Sing, D.~K., Goyal, J.~M., {et~al.} 2019, Monthly Notices of the Royal Astronomical Society, 488, 2222

\bibitem[{Molaverdikhani {et~al.}(2019)Molaverdikhani, Henning, \& Mollière}]{molaverdikhani_cold_2019}
Molaverdikhani, K., Henning, T., \& Mollière, P. 2019, The Astrophysical Journal, 883, 194

\bibitem[{Moses(2014)}]{moses_chemical_2014}
Moses, J.~I. 2014, Philosophical Transactions of the Royal Society A: Mathematical, Physical and Engineering Sciences, 372, 20130073

\bibitem[{Moses {et~al.}(2011)Moses, Visscher, Fortney, Showman, Lewis, Griffith, Klippenstein, Shabram, Friedson, Marley, \& Freedman}]{moses_disequilibrium_2011}
Moses, J.~I., Visscher, C., Fortney, J.~J., {et~al.} 2011, The Astrophysical Journal, 737, 15

\bibitem[{Nortmann {et~al.}(2024)Nortmann, Lesjak, Yan, Cont, Czesla, Lavail, Rains, Nagel, Boldt-Christmas, Hatzes, Reiners, Piskunov, Kochukhov, Heiter, Shulyak, Rengel, \& Seemann}]{nortmann_crires_2024}
Nortmann, L., Lesjak, F., Yan, F., {et~al.} 2024, {CRIRES}{\textbackslash}ˆ+{\textbackslash} transmission spectroscopy of {WASP}-127b. {Detection} of the resolved signatures of a supersonic equatorial jet and cool poles in a hot planet, publication Title: arXiv.org

\bibitem[{Nugroho {et~al.}(2021)Nugroho, Kawahara, Gibson, Mooij, Hirano, Kotani, Kawashima, Masuda, Brogi, Birkby, Watson, Tamura, Zwintz, Harakawa, Kudo, Kuzuhara, Hodapp, Ishizuka, Jacobson, Konishi, Kurokawa, Nishikawa, Omiya, Serizawa, Ueda, \& Vievard}]{nugroho_first_2021}
Nugroho, S.~K., Kawahara, H., Gibson, N.~P., {et~al.} 2021, The Astrophysical Journal Letters, 910, L9

\bibitem[{Oliva {et~al.}(2015)Oliva, Origlia, Scuderi, Benatti, Carleo, Lapenna, Mucciarelli, Baffa, Biliotti, Carbonaro, Falcini, Giani, Iuzzolino, Massi, Sanna, Sozzi, Tozzi, Ghedina, Ghinassi, Lodi, Harutyunyan, \& Pedani}]{oliva_lines_2015}
Oliva, E., Origlia, L., Scuderi, S., {et~al.} 2015, Astronomy \& Astrophysics, 581, A47

\bibitem[{Ouyang {et~al.}(2023)Ouyang, Wang, Zhai, Chen, Rojo, Liu, Zhao, Huang, \& Zhao}]{ouyang_detection_2023}
Ouyang, Q., Wang, W., Zhai, M., {et~al.} 2023, Research in Astronomy and Astrophysics, 23, 065010

\bibitem[{Parmentier \& Crossfield(2018)}]{parmentier_exoplanet_2018}
Parmentier, V. \& Crossfield, I. J.~M. 2018, Exoplanet {Phase} {Curves}: {Observations} and {Theory}, pages: 116 Publication Title: Handbook of Exoplanets ADS Bibcode: 2018haex.bookE.116P

\bibitem[{Parmentier {et~al.}(2018)Parmentier, Line, Bean, Mansfield, Kreidberg, Lupu, Visscher, Désert, Fortney, Deleuil, Arcangeli, Showman, \& Marley}]{parmentier_thermal_2018}
Parmentier, V., Line, M.~R., Bean, J.~L., {et~al.} 2018, Astronomy \& Astrophysics, 617, A110

\bibitem[{Pelletier {et~al.}(2023)Pelletier, Benneke, Ali-Dib, Prinoth, Kasper, Seifahrt, Bean, Debras, Klein, Bazinet, Hoeijmakers, Kesseli, Lim, Carmona, Pino, Casasayas-Barris, Hood, \& Stürmer}]{pelletier_vanadium_2023}
Pelletier, S., Benneke, B., Ali-Dib, M., {et~al.} 2023, Nature, 619, 491

\bibitem[{Pelletier {et~al.}(2024)Pelletier, Benneke, Chachan, Bazinet, Allart, Hoeijmakers, Lavail, Prinoth, Coulombe, Lothringer, Parmentier, Smith, Borsato, \& Thorsbro}]{pelletier_crires_2024}
Pelletier, S., Benneke, B., Chachan, Y., {et~al.} 2024, The Astronomical Journal, 169, 10, publisher: The American Astronomical Society

\bibitem[{Pelletier {et~al.}(2021)Pelletier, Benneke, Darveau-Bernier, Boucher, Cook, Piaulet, Coulombe, Artigau, Lafrenière, Delisle, Allart, Doyon, Donati, Fouqué, Moutou, Cadieux, Delfosse, Hébrard, Martins, Martioli, \& Vandal}]{pelletier_where_2021}
Pelletier, S., Benneke, B., Darveau-Bernier, A., {et~al.} 2021, The Astronomical Journal, 162, 73

\bibitem[{Pepe {et~al.}(2021)Pepe, Cristiani, Rebolo, Santos, Dekker, Cabral, Di~Marcantonio, Figueira, Lo~Curto, Lovis, Mayor, Mégevand, Molaro, Riva, Zapatero~Osorio, Amate, Manescau, Pasquini, Zerbi, Adibekyan, Abreu, Affolter, Alibert, Aliverti, Allart, Allende~Prieto, Álvarez, Alves, Avila, Baldini, Bandy, Barros, Benz, Bianco, Borsa, Bourrier, Bouchy, Broeg, Calderone, Cirami, Coelho, Conconi, Coretti, Cumani, Cupani, D'Odorico, Damasso, Deiries, Delabre, Demangeon, Dumusque, Ehrenreich, Faria, Fragoso, Genolet, Genoni, Génova~Santos, González~Hernández, Hughes, Iwert, Kerber, Knudstrup, Landoni, Lavie, Lillo-Box, Lizon, Maire, Martins, Mehner, Micela, Modigliani, Monteiro, Monteiro, Moschetti, Murphy, Nunes, Oggioni, Oliveira, Oshagh, Pallé, Pariani, Poretti, Rasilla, Rebordão, Redaelli, Santana~Tschudi, Santin, Santos, Ségransan, Schmidt, Segovia, Sosnowska, Sozzetti, Sousa, Spanò, Suárez~Mascareño, Tabernero, Tenegi, Udry, \& Zanutta}]{pepe_espresso_2021}
Pepe, F., Cristiani, S., Rebolo, R., {et~al.} 2021, Astronomy and Astrophysics, 645, A96

\bibitem[{Pepe {et~al.}(2002)Pepe, Mayor, Rupprecht, Avila, Ballester, Beckers, Benz, Bertaux, Bouchy, Buzzoni, Cavadore, Deiries, Dekker, Delabre, D'Odorico, Eckert, Fischer, Fleury, George, Gilliotte, Gojak, Guzman, Koch, Kohler, Kotzlowski, Lacroix, Le~Merrer, Lizon, Lo~Curto, Longinotti, Megevand, Pasquini, Petitpas, Pichard, Queloz, Reyes, Richaud, Sivan, Sosnowska, Soto, Udry, Ureta, van Kesteren, Weber, Weilenmann, Wicenec, Wieland, Christensen-Dalsgaard, Dravins, Hatzes, Kürster, Paresce, \& Penny}]{pepe_harps_2002}
Pepe, F., Mayor, M., Rupprecht, G., {et~al.} 2002, The Messenger, 110, 9, aDS Bibcode: 2002Msngr.110....9P

\bibitem[{Polanski {et~al.}(2022)Polanski, Crossfield, Howard, Isaacson, \& Rice}]{polanski_chemical_2022}
Polanski, A.~S., Crossfield, I. J.~M., Howard, A.~W., Isaacson, H., \& Rice, M. 2022, Research Notes of the AAS, 6, 155

\bibitem[{Polyansky {et~al.}(2018)Polyansky, Kyuberis, Zobov, Tennyson, Yurchenko, \& Lodi}]{polyansky_exomol_2018}
Polyansky, O.~L., Kyuberis, A.~A., Zobov, N.~F., {et~al.} 2018, Monthly Notices of the Royal Astronomical Society, 480, 2597

\bibitem[{Prinoth {et~al.}(2022)Prinoth, Hoeijmakers, Kitzmann, Sandvik, Seidel, Lendl, Borsato, Thorsbro, Anderson, Barrado, Kravchenko, Allart, Bourrier, Cegla, Ehrenreich, Fisher, Lovis, Guzmán-Mesa, Grimm, Hooton, Morris, Oreshenko, Pino, \& Heng}]{prinoth_titanium_2022}
Prinoth, B., Hoeijmakers, H.~J., Kitzmann, D., {et~al.} 2022, Nature Astronomy, 6, 449

\bibitem[{Rauscher \& Menou(2012)}]{rauscher_general_2012}
Rauscher, E. \& Menou, K. 2012, The Astrophysical Journal, 750, 96

\bibitem[{Rauscher \& Menou(2013)}]{rauscher_three-dimensional_2013}
Rauscher, E. \& Menou, K. 2013, The Astrophysical Journal, 764, 103

\bibitem[{Roth {et~al.}(2021)Roth, Drummond, Hébrard, Tremblin, Goyal, \& Mayne}]{roth_pseudo-2d_2021}
Roth, A., Drummond, B., Hébrard, E., {et~al.} 2021, Monthly Notices of the Royal Astronomical Society, 505, 4515

\bibitem[{Rothman {et~al.}(2010)Rothman, Gordon, Barber, Dothe, Gamache, Goldman, Perevalov, Tashkun, \& Tennyson}]{rothman_hitemp_2010}
Rothman, L.~S., Gordon, I.~E., Barber, R.~J., {et~al.} 2010, Journal of Quantitative Spectroscopy and Radiative Transfer, 111, 2139

\bibitem[{Roudier {et~al.}(2021)Roudier, Swain, Gudipati, West, Estrela, \& Zellem}]{roudier_disequilibrium_2021}
Roudier, G.~M., Swain, M.~R., Gudipati, M.~S., {et~al.} 2021, The Astronomical Journal, 162, 37

\bibitem[{Ryabchikova {et~al.}(2015)Ryabchikova, Piskunov, Kurucz, Stempels, Heiter, Pakhomov, \& Barklem}]{ryabchikova_major_2015}
Ryabchikova, T., Piskunov, N., Kurucz, R.~L., {et~al.} 2015, Physica Scripta, 90, 054005

\bibitem[{Seidel {et~al.}(2023)Seidel, Borsa, Pino, Ehrenreich, Stangret, Osorio, Palle, Alibert, Allart, Bourrier, Marcantonio, Figueira, Hernández, Lillo-Box, Lovis, Martins, Mehner, Molaro, Nunes, Pepe, Santos, \& Sozzetti}]{seidel_detection_2023}
Seidel, J.~V., Borsa, F., Pino, L., {et~al.} 2023, Astronomy \& Astrophysics, 673, A125

\bibitem[{Seidel {et~al.}(2021)Seidel, Ehrenreich, Allart, Hoeijmakers, Lovis, Bourrier, Pino, Wyttenbach, Adibekyan, Alibert, Borsa, Casasayas-Barris, Cristiani, Demangeon, Marcantonio, Figueira, Hernández, Lillo-Box, Martins, Mehner, Molaro, Nunes, Palle, Pepe, Santos, Sousa, Sozzetti, Tabernero, \& Osorio}]{seidel_into_2021}
Seidel, J.~V., Ehrenreich, D., Allart, R., {et~al.} 2021, Astronomy \& Astrophysics, 653, A73

\bibitem[{Showman {et~al.}(2009)Showman, Fortney, Lian, Marley, Freedman, Knutson, \& Charbonneau}]{showman_atmospheric_2009}
Showman, A.~P., Fortney, J.~J., Lian, Y., {et~al.} 2009, The Astrophysical Journal, 699, 564

\bibitem[{Smith {et~al.}(2024)Smith, Sanchez, Line, Rauscher, Mansfield, Kempton, Savel, Wardenier, Pino, Bean, Beltz, Panwar, Brogi, Malsky, Fortney, Désert, Pelletier, Parmentier, Kanumalla, Welbanks, Meyer, \& Monnier}]{smith_roasting_2024}
Smith, P. C.~B., Sanchez, J.~A., Line, M.~R., {et~al.} 2024, The Astronomical Journal, 168, 293, publisher: The American Astronomical Society

\bibitem[{Snellen {et~al.}(2010)Snellen, de~Kok, de~Mooij, \& Albrecht}]{snellen_orbital_2010}
Snellen, I. A.~G., de~Kok, R.~J., de~Mooij, E. J.~W., \& Albrecht, S. 2010, Nature, 465, 1049

\bibitem[{Stock {et~al.}(2022)Stock, Kitzmann, \& Patzer}]{stock_fastchem_2022}
Stock, J.~W., Kitzmann, D., \& Patzer, A. B.~C. 2022, Monthly Notices of the Royal Astronomical Society, 517, 4070

\bibitem[{Stock {et~al.}(2018)Stock, Kitzmann, Patzer, \& Sedlmayr}]{stock_fastchem_2018}
Stock, J.~W., Kitzmann, D., Patzer, A. B.~C., \& Sedlmayr, E. 2018, Monthly Notices of the Royal Astronomical Society, 479, 865

\bibitem[{Tan \& Komacek(2019)}]{tan_atmospheric_2019}
Tan, X. \& Komacek, T.~D. 2019, The Astrophysical Journal, 886, 26

\bibitem[{Tan {et~al.}(2024)Tan, Komacek, Batalha, Deming, Lupu, Parmentier, \& Pierrehumbert}]{tan_modelling_2024}
Tan, X., Komacek, T.~D., Batalha, N.~E., {et~al.} 2024, Monthly Notices of the Royal Astronomical Society, 528, 1016

\bibitem[{Thiabaud {et~al.}(2015)Thiabaud, Marboeuf, Alibert, Leya, \& Mezger}]{thiabaud_elemental_2015}
Thiabaud, A., Marboeuf, U., Alibert, Y., Leya, I., \& Mezger, K. 2015, Astronomy and Astrophysics, 580, A30, aDS Bibcode: 2015A\&A...580A..30T

\bibitem[{Thorngren \& Fortney(2019)}]{thorngren_connecting_2019}
Thorngren, D. \& Fortney, J.~J. 2019, The Astrophysical Journal Letters, 874, L31

\bibitem[{van Sluijs {et~al.}(2023)van Sluijs, Birkby, Lothringer, Lee, Crossfield, Parmentier, Brogi, Kulesa, McCarthy, \& Charbonneau}]{van_sluijs_carbon_2023}
van Sluijs, L., Birkby, J.~L., Lothringer, J., {et~al.} 2023, Monthly Notices of the Royal Astronomical Society, 522, 2145

\bibitem[{Wardenier {et~al.}(2025)Wardenier, Parmentier, Lee, \& Line}]{wardenier_pretransit_2025}
Wardenier, J.~P., Parmentier, V., Lee, E. K.~H., \& Line, M.~R. 2025, The Astrophysical Journal, 986, 63, publisher: The American Astronomical Society

\bibitem[{Wardenier {et~al.}(2021)Wardenier, Parmentier, Lee, Line, \& Gharib-Nezhad}]{wardenier_decomposing_2021}
Wardenier, J.~P., Parmentier, V., Lee, E. K.~H., Line, M.~R., \& Gharib-Nezhad, E. 2021, Monthly Notices of the Royal Astronomical Society, 506, 1258

\bibitem[{Wardenier {et~al.}(2023)Wardenier, Parmentier, Line, \& Lee}]{wardenier_modelling_2023}
Wardenier, J.~P., Parmentier, V., Line, M.~R., \& Lee, E. K.~H. 2023, Monthly Notices of the Royal Astronomical Society, 525, 4942

\bibitem[{Wardenier {et~al.}(2024)Wardenier, Parmentier, Line, Mansfield, Tan, Tsai, Bean, Birkby, Brogi, Désert, Gandhi, Lee, Levens, Pino, \& Smith}]{wardenier_phase-resolving_2024}
Wardenier, J.~P., Parmentier, V., Line, M.~R., {et~al.} 2024, Phase-resolving the absorption signatures of water and carbon monoxide in the atmosphere of the ultra-hot {Jupiter} {WASP}-121b with {GEMINI}-{S}/{IGRINS}

\bibitem[{West {et~al.}(2016)West, Hellier, Almenara, Anderson, Barros, Bouchy, Brown, Cameron, Deleuil, Delrez, Doyle, Faedi, Fumel, Gillon, Chew, Hébrard, Jehin, Lendl, Maxted, Pepe, Pollacco, Queloz, Ségransan, Smalley, Smith, Southworth, Triaud, \& Udry}]{west_three_2016}
West, R.~G., Hellier, C., Almenara, J.-M., {et~al.} 2016, Astronomy \& Astrophysics, 585, A126, publisher: EDP Sciences

\bibitem[{Wildi {et~al.}(2022)Wildi, Bouchy, Doyon, Blind, Genolet, Sordet, Segovia, Grieves, Malo, Artigau, St-Antoine, Vallée, Rasilla, Gracia~Temich, Poulin-Girard, Brousseau, Sosnowska, Reshetov, Baron, Thibault, Bovay, Frensch, Lo~Curto, Hubin, Zins, Peroux, \& Cabral}]{wildi_first_2022}
Wildi, F., Bouchy, F., Doyon, R., {et~al.} 2022, 12184, 121841H, conference Name: Ground-based and Airborne Instrumentation for Astronomy IX ADS Bibcode: 2022SPIE12184E..1HW

\bibitem[{Yurchenko {et~al.}(2020)Yurchenko, Mellor, Freedman, \& Tennyson}]{yurchenko_exomol_2020}
Yurchenko, S.~N., Mellor, T.~M., Freedman, R.~S., \& Tennyson, J. 2020, Monthly Notices of the Royal Astronomical Society, 496, 5282

\bibitem[{Yurchenko {et~al.}(2022)Yurchenko, Tennyson, Syme, Adam, Clark, Cooper, Dobney, Donnelly, Gorman, Lynas-Gray, Meltzer, Owens, Qu, Semenov, Somogyi, Upadhyay, Wright, \& Zapata~Trujillo}]{yurchenko_exomol_2022}
Yurchenko, S.~N., Tennyson, J., Syme, A.-M., {et~al.} 2022, Monthly Notices of the Royal Astronomical Society, 510, 903

\end{thebibliography}

\begin{appendix}
\onecolumn

\section{Comparison between pre-processing pipelines}

\begin{figure*}
    \centering
    \includegraphics[width=\linewidth]{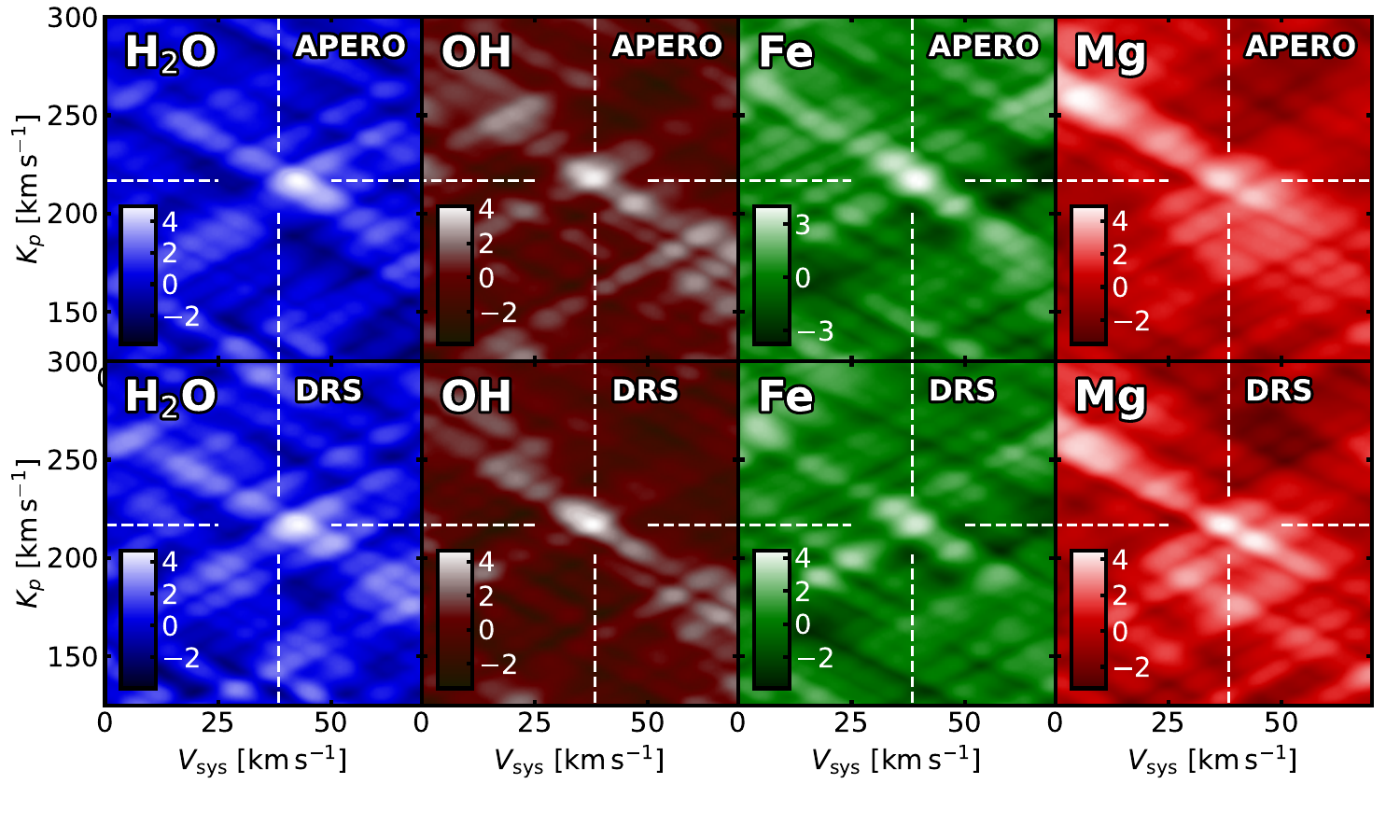}
    \caption{Comparison of the cross-correlation maps between both pre-processing pipelines. The top panels use data pre-processed by APERO (Same as Fig. \ref{fig:kpvsys}). The bottom panels use the NIRPS-DRS pipeline.}
    \label{fig:apero_vs_geneva}
\end{figure*}

NIRPS has two data reduction pipelines: APERO and NIRPS-DRS. Figure \ref{fig:apero_vs_geneva} shows the cross-correlation maps of the four detected species, where the only difference in the data reduction steps is the choice of the pre-processing pipeline. The maps are similar. All features, such as the relative shifts, are present regardless of the pipeline used. Minor differences between maps can be attributed by the different approaches to the telluric correction.

\section{Principal components removal}

\begin{figure}
    \centering
    \includegraphics[width=0.5\linewidth]{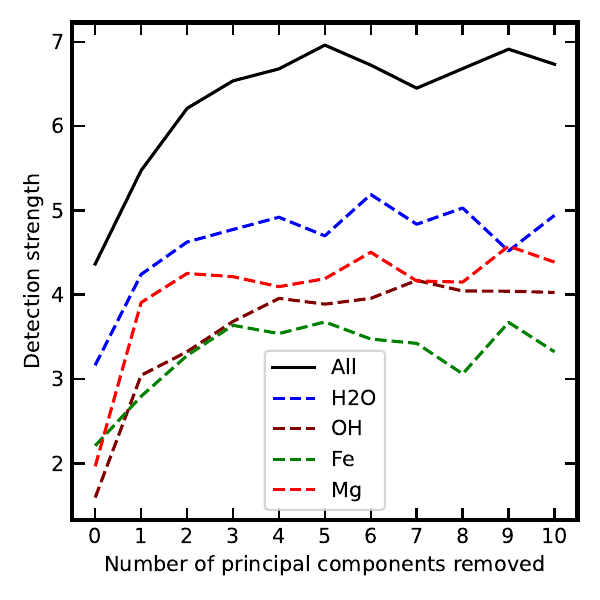}
    \caption{Evolution of signals strengths as a function of the number of principal components. Signal strengths are defined as the maximum in the cross-correlation signal-to-noise $K_p$-$V_{\mathrm{sys}}$ map in a 12 km\,s$^{-1}$ by 12 km\,s$^{-1}$ region centred on the expected velocity. 
    This is done for an atmospheric model that includes all four species: H$_2$O, OH, Fe and Mg (black solid line) as well as separately for each species (coloured dashed lines).}
    \label{fig:detect_vs_PCA}
\end{figure}

A crucial step in our data detrending procedure is the removal of the first few principal components (PCs). The choice of how many PCs should be removed from a given data set has been the subject of several studies \citep[e.g.,][]{cheverall_robustness_2023}. Typically, a few PCs are needed to remove the dominant contribution from tellurics, however removing too many components can risk removing the planetary signal of interest. Here, similar as in \cite{holmberg_first_2022}, we explore the evolution of the observed planetary signal as a function of the number of PCs removed (Fig. \ref{fig:detect_vs_PCA}). All models exhibit the same general trend. When the number of components removed is low, the strength is weak. The strength increases as the number of PCs removed increases, up to a certain point where the strength remain somewhat constant. This point varies for one model to another, however all of them become constant after $\sim$4 PCs.

\begin{figure}
    \centering
    \includegraphics[width=0.9\linewidth]{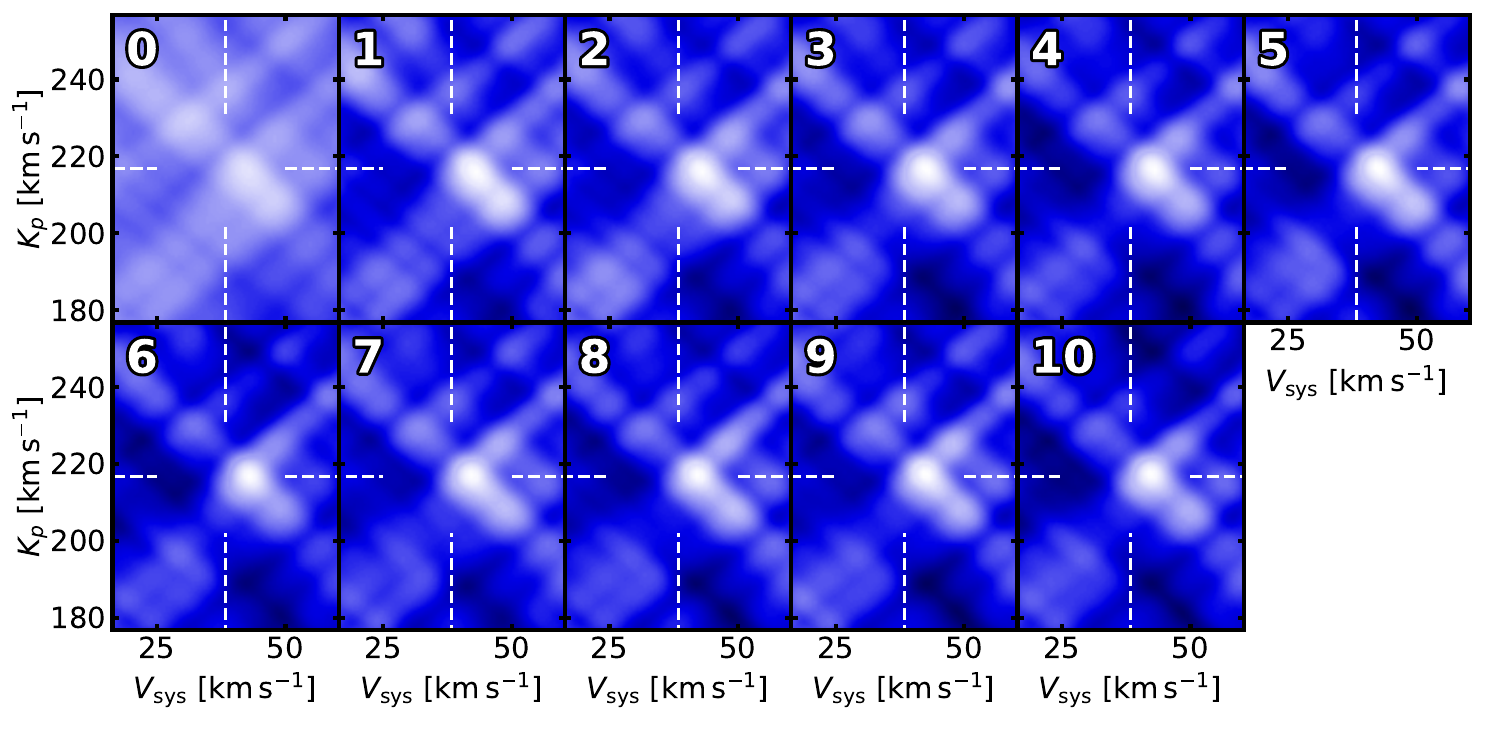}
    \caption{Cross-correlation signal-to-noise maps of H$_2$O where we vary the number of principal components removed (top left number of each panel).  The H$_2$O signal position does not change significantly based on the number of components removed.}
    \label{fig:H2O_Kp_Vsys_vs_PCA}
\end{figure}

To determine whether the H$_2$O shift is dependent on the reduction steps, we also generate $K_p$-$V_\mathrm{sys}$ signal-to-noise maps of H$_2$O varying the number of PCs removed (Fig. \ref{fig:H2O_Kp_Vsys_vs_PCA}). The position of the H$_2$O signal stays between $V_{\mathrm{sys}}$ of 42 and 43 km\,s$^{-1}$, regardless of the number of PCs removed.  For our main analysis, we choose to remove 5 PCs. However, we also run a retrieval where the only difference is that we set the number of components removed to 10. The results are consistent in both cases (Table \ref{tab:results_5_10_pcs}).

\begin{table}[]
    \centering
    \renewcommand{\arraystretch}{1.35}
    \begin{tabular}{c c c}
    \hline
     Parameter [unit] & 5 PCs & 10 PCs \\
     \hline
     log$_{10}$H$_2$O & $-3.49^{+0.99}_{-0.93}$ &  $-3.49^{+0.99}_{-1.34}$\\
     log$_{10}$OH & $-3.59^{+1.01}_{-1.08}$ & $-3.62^{+1.00}_{-1.35}$\\
     log$_{10}$Fe & $-2.91^{+1.10}_{-1.36}$ & $-3.19^{+1.13}_{-1.57}$\\
     log$_{10}$Mg & $-2.75^{+1.09}_{-1.37}$ & $-3.01^{+1.18}_{-1.41}$\\
     log$_{10}$H$^{-}$ & $< -6.06$ (3$\,\sigma$ limit) & $< -6.28$ (3$\,\sigma$ limit) \\
     log$_{10}$e$^{-}$ & $< -1.03$ (3$\,\sigma$ limit) & $< -1.04$ (3$\,\sigma$ limit) \\
     $K_{p}$ [km\,s$^{-1}$] & $217.21^{+0.63}_{-0.62}$ & $217.01^{+0.58}_{-0.66}$\\
     $V_{\mathrm{sys}}$ [km\,s$^{-1}$] & $38.04\pm{0.52}$ & $38.12^{+0.44}_{-0.56}$\\
     $\Delta V_{\mathrm{sys, H_2O}}$ [km\,s$^{-1}$] & $4.79^{+0.93}_{-0.97}$ & $4.24\pm{0.91}$\\
     \hline
    \end{tabular}
    \caption{Results of the retrieval for the retrievals where 5 principal components (5 PCs) were removed (same as the main retrieval presented in this article) and where 10 principal components (10 PCs) were removed.}
    \label{tab:results_5_10_pcs}
\end{table}

\section{Pre- and post- phases cross-correlation maps}
\begin{figure*}
    \centering
    \includegraphics[width=\linewidth]{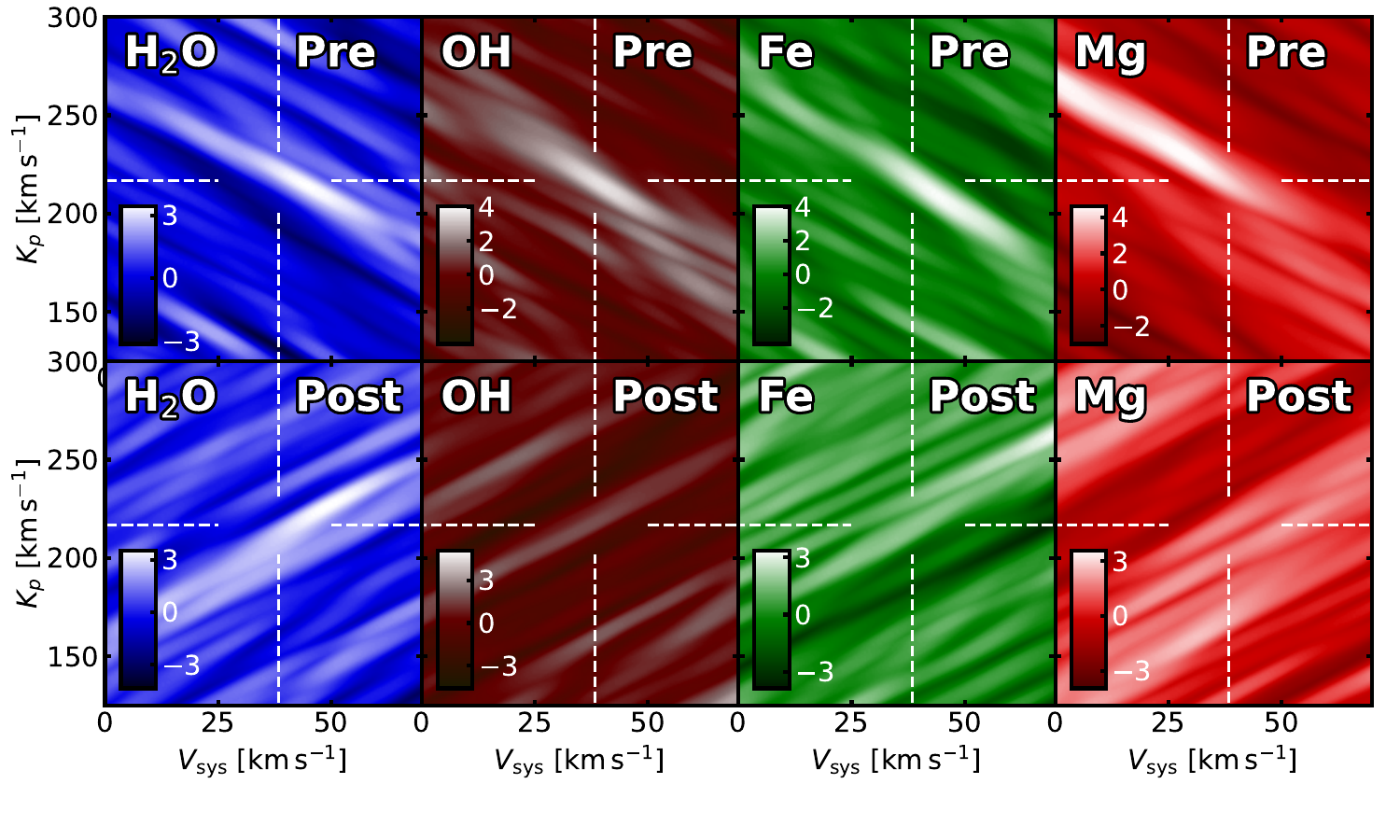}
    \caption{Same as Fig. \ref{fig:kpvsys} but where the top row is the map only considering the pre-eclipse phases observations and the bottom map is only considering the post-eclipse phases.}
    \label{fig:kp_vsys_pre_post}
\end{figure*}

Figure \ref{fig:kp_vsys_pre_post} shows the cross-correlation maps where we only consider the pre-eclipse phases observation or only consider the post-eclipse phases observations. Overall, the pre-eclipse phases seem to depict better detections.

\section{GCM cross-correlation maps}

\begin{figure*}
    \centering
    \includegraphics[width=\linewidth]{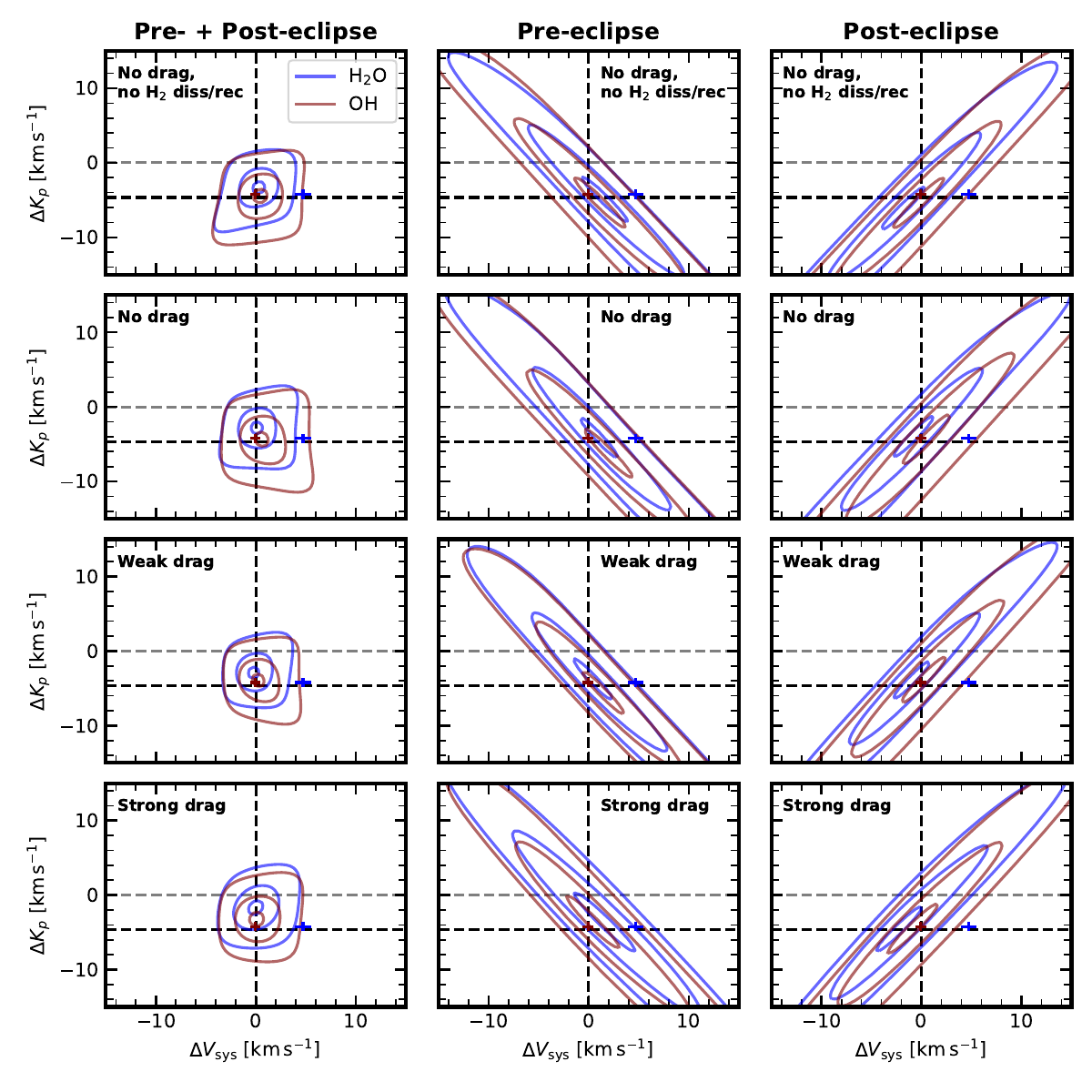}
    \caption{Same as Fig. \ref{fig:GCM_strongdrag} but for four different GMCs.
    The top row has models that has no drag and does not consider latent heat release from dissociation and recombination of H$_2$ \citep{parmentier_thermal_2018}. The next three rows are maps of models that considers the dissociation/recombination of H$_2$ \citep{tan_modelling_2024}. The difference comes in the drag timescale: $\tau_{\mathrm{drag}}=\infty$ (no drag), $\tau_{\mathrm{drag}}=10^6$\,s (weak drag), and $\tau_{\mathrm{drag}}=10^4$\,s (strong drag).
    The left column maps consider both pre- and post-eclipse phases. 
    The middle column includes the maps when considering only the pre-eclipse phases. The right column are the maps of the post-eclipse phases only.
    }
    \label{fig:GCMs_4_models}
\end{figure*}

Figure \ref{fig:GCMs_4_models} shows the cross-correlation maps of the four considered GCMs, along with the detection locations. All models predict the decrease of the expected $K_p$ by the synchronous rotation of the planet.

\end{appendix}

\end{document}